\documentclass[twocolumn,10pt]{IEEEtran}
\usepackage{verbatim}
\usepackage{epsfig,bbm}
\usepackage[colorlinks,bookmarksopen,bookmarksnumbered,citecolor=red,urlcolor=red,]{hyperref}
\usepackage{CJK}
\usepackage{indentfirst}
\usepackage{multirow}
\usepackage{epstopdf}
\usepackage{graphicx}
\usepackage{footmisc}
\graphicspath{{./figures/}}
\usepackage{amsfonts}
\usepackage{mathrsfs}
\usepackage{setspace}
\usepackage{amsmath}
\usepackage{algorithm,algorithmic,amsbsy,amsmath,amssymb,epsfig,bbm,mathrsfs, bbm} 
\usepackage{amsthm}
\usepackage{verbatim}
\usepackage[noadjust]{cite}
\usepackage[subfigure]{graphfig}

\begin{document}

\title{Wireless Networks with RF Energy Harvesting: A Contemporary Survey}
 \author{   Xiao Lu$^{\dagger}$, Ping Wang$^{\dagger}$,  Dusit Niyato$^{\dagger}$, Dong In Kim$^{\ddagger}$, and Zhu Han$^{\S}$\\
    ~$^{\dagger}$ School of Computer Engineering, Nanyang Technological University, Singapore \\
    $^{\ddagger}$  School of Information and Communication     Engineering, Sungkyunkwan University (SKKU), Korea \\
    $^{\S}$  Electrical and Computer Engineering, University of Houston, Texas, USA.
    \thanks{\textbf{X. Lu} is the corresponding author. Any comment(s) would be highly welcomed. Please send email to luxiao@ntu.edu.sg}
 }    
\maketitle


\begin{abstract}
 
Radio frequency (RF) energy transfer and harvesting techniques have recently become alternative methods to power the next generation wireless networks. As this emerging technology enables proactive energy replenishment of wireless devices, it is advantageous in supporting applications with quality of service (QoS) requirements. In this paper, we present an extensive literature review on the research progresses in wireless networks with RF energy harvesting capability, referred to as RF energy harvesting networks (RF-EHNs). First, we present an overview of the RF-EHNs including system architecture, RF energy harvesting techniques and existing applications. Then, we present the background in circuit design as well as the state-of-the-art circuitry implementations, and review the communication protocols specially designed for RF-EHNs. We also explore various key design issues in the development of RF-EHNs according to the network types, i.e., single-hop networks, multi-antenna networks, relay networks, and cognitive radio networks. Finally, we envision some open research directions.

\end{abstract}

\emph{Index terms- RF energy harvesting, simultaneous wireless information and power transfer (SWIPT), receiver operation policy, beamforming, communication protocols, RF-powered Cognitive radio network}. 

\IEEEpeerreviewmaketitle
\section{Introduction}
 
Recently, there has been an upsurge of research interests in radio frequency (RF) energy harvesting/scavenging technique (see~\cite{Visser2013} and references therein), or RF harvesting in short, which is the capability of converting the received RF signals into electricity. This technique becomes a promising solution to power energy-constrained wireless networks. Conventionally, the energy-constrained wireless networks, such as wireless sensor networks, have a limited lifetime which largely confines the network performance. In contrast, an RF energy harvesting network (RF-EHN) has a sustainable power supply from a radio environment. Therefore, the RF energy harvesting capability allows the wireless devices to harvest energy from RF signals for their information processing and transmission. Consequently, RF-EHNs have found their applications quickly in various forms, such as wireless sensor networks~\cite{Nishimoto2010}, wireless body networks~\cite{Zhang2010}, and wireless charging systems. With the increasingly emerging applications of RF energy harvesting/charging, the Wireless Power Consortium is also making the efforts of establishing an international standard for the RF energy harvesting technique.

In RF energy harvesting, radio signals with frequency range from 300GHz  to as low as 3kHz are used as a medium to carry energy in a form of electromagnetic radiation. RF energy transfer and harvesting is one of the wireless energy transfer techniques. The other techniques are inductive coupling and magnetic resonance coupling. Inductive coupling~\cite{Liu2011Henry} is based on magnetic coupling that delivers electrical energy between two coils tuned to resonate at the same frequency. The electric power is carried through the magnetic field between two coils. Magnetic resonance coupling~\cite{Kurs2007A} utilizes evanescent-wave coupling to generate and transfer electrical energy between two resonators. The resonator is formed by adding a capacitance on an induction coil. Both of the above two techniques are near-field wireless transmission featured with high power density and conversion efficiency. The power transmission efficiency depends on the coupling coefficient, which depends on the distance between two coils/resonators. The power strength is attenuated according to the cube of the reciprocal of the distance \cite{J.2010O,Tutorial}, specifically, 60 dB per decade of the distance, which results in limited power transfer distance. Besides, both inductive coupling and resonance coupling require calibration and alignment of coils/resonators at transmitters and receivers. Therefore, they are not suitable for mobile and remote replenishment/charging. In contrast, RF energy transfer has no such limitation. As the radiative electromagnetic wave cannot retroact upon the antenna that generated it (by capacitive or inductive coupling) at a distance of above $\lambda /(2 \pi)$ \cite{C1973Johnson}, RF energy transfer can be regarded as a far-field energy transfer technique. Thus, RF energy transfer is suitable for powering a larger number of devices distributed in a wide area. The signal strength of far-field RF transmission is attenuated according to the reciprocal of the distance between transmitter and receiver, specifically, 20 dB  per decade of the distance. Table~\ref{WET} shows the comparison between the three major wireless energy transfer techniques. We can see that RF energy transfer technique has clear advantages in effective energy transfer distance. However, it has low RF-to-DC energy conversion efficiency especially when the harvested RF power is small. The readers can refer to \cite{N2012Shinohara,L2013Xie} for more detailed introduction of wireless energy transfer techniques. In this article, we focus on wireless networks with the RF energy harvesting technique.

Wireless power transfer has caught research attention since long ago, as a separate problem with wireless information transmission. Traditionally, free-space beaming and antennas with large apertures were used to overcome propagation loss for large power transfer. For example, in 1960's, the authors in \cite{C1969Brown} demonstrate a small helicopter hovering at an height of 50-feet, powered by an RF source with a DC power supply of 270W operating on 2.45GHz on the ground. In \cite{O2002Mcspadden}, the authors demonstrate a space-to-earth power transfer system using gigantic transmit antenna arrays at a satellite and receive antenna arrays at a ground station. For transmit power of 2.7GW, the power transfer efficiency is estimated to be $45\%$ over a transfer distance of $36000km$. During the past decade, with the development in RF energy harvesting circuit, low power transfer for powering mobile terminals in wireless communication systems began to attract increasing attention \cite{K1207.5640Huang,L1312.1450Liu}. The authors in \cite{K1207.5640Huang} propose a network architecture for RF charging stations, overlaying with an uplink cellular network. In \cite{L1312.1450Liu}, a harvest-then-transmit protocol is introduced for power transfer in wireless broadcast system. Moreover, various modern beamforming techniques are employed to improve power transfer efficiency \cite{L1312.1450Liu,G1311.4111Yang,X2013Chen} for mobile applications.

\begin{table*}
\footnotesize
\centering
\caption{\footnotesize Comparison of different wireless energy transfer techniques.} \label{WET}
\begin{tabular}{|p{2.5cm}|p{1.5cm}|p{1.6cm}|p{3.5cm}|p{3cm}|p{3cm}|} 
\hline
\footnotesize {\bf Wireless energy transfer technique} & {\bf Field region} & {\bf Propagation} & {\bf Effective distance} & {\bf Efficiency} & {\bf Applications} \\
\hline
RF energy transfer & Far-field & Radiative & Depend on distance and frequency and the sensitivity of RF energy harvester (typically from several meters to several kilometers) & $0.4\%$, above $18.2\%$, and over $50\%$ at $-40$ dBm, -$20$ dbm and -$5$ dBm input power, respectively~\cite{C2011Mikeka} & Wireless sensor network~\cite{Nishimoto2010}, wireless body network~\cite{Zhang2010} \\
\hline
Resonant inductive coupling& Near-field & Non-radiative & From a few millimeters to a few centimeters & From $5.81\%$ to $57.2\%$ when frequency varies from 16.2kHz to 508kHz~\cite{Liu2011Henry} & Passive
RF indentification (RFID) tags, contactless smart cards, cell phone charging \\
\hline
Magnetic resonance coupling & Near-field & Non-radiative &   From a few centimeters to a few meters & From above $90\%$ to above $30\%$ when distance varies from 0.75m to 2.25m~\cite{Kurs2007A} & PHEV charging, cell phone charging \\ 
\hline
\end{tabular}
\end{table*}

It is until recently that the dual use of RF signals for delivering energy as well as for transporting information has been advocated~\cite{Varshney2008,X2014Lu}. Simultaneous wireless information and power transfer (SWIPT) \cite{ZhangRuiMIMO} is proposed for delivering RF energy, usually in a low power region (e.g., for sensor networks). SWIPT provides the advantage of delivering controllable and efficient on-demand wireless information and energy concurrently, which offers a low-cost option for sustainable operations of wireless systems without hardware modification on the transmitter side. However, recent research has recognized that optimizing wireless information and energy transfer simultaneously brings tradeoff on the design of a wireless system \cite{Varshney2008,P2010Grover}. The reason can be understood as the amount of ``variations", i.e., entropy rate, in an RF signal determines the quantity of information, while the average squared value of RF signals account for its power. Consequently, the amount of transmitted information and transferred energy cannot be generally maximized at the same time. This raises a demand for redesign of existing wireless networks.

\begin{figure}
\centering
\includegraphics[width=0.45\textwidth]{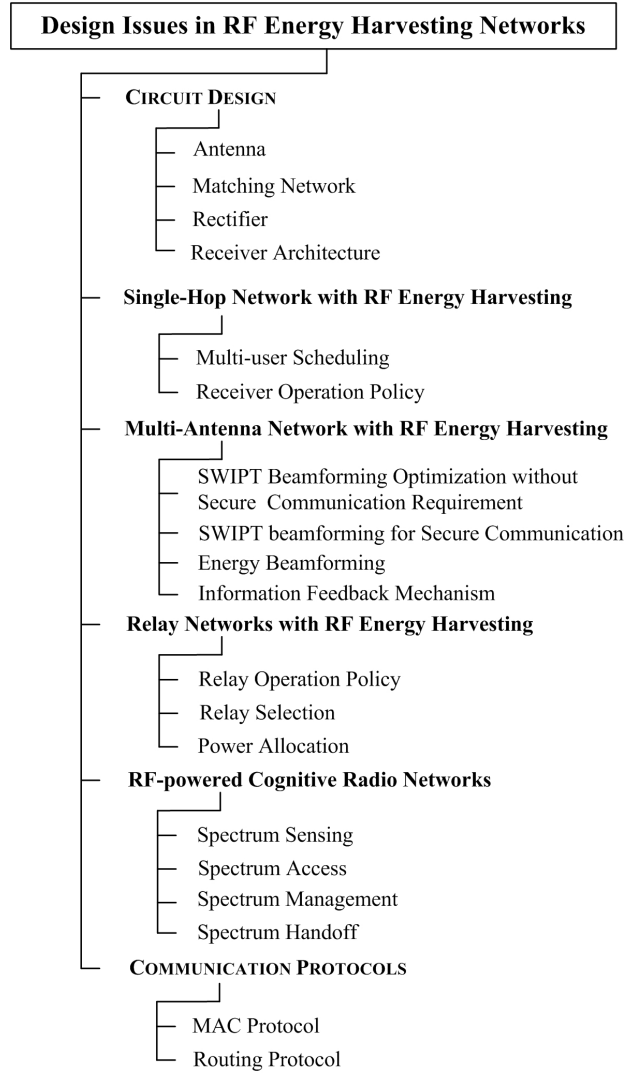}
\caption{Outline of design issues in RF-EHNs} \label{outline}
\end{figure}

This paper aims to provide a comprehensive survey on the contemporary research in RF-EHNs. The scope of this survey covers circuit design, communication protocols as well as the emerging operation designs in various types of RF-EHNs. Note that we emphasize on the design issues related to communications in RF-EHNs. The hardware technology for RF energy harvesting electronics is beyond the scope of this paper. Figure~\ref{outline} outlines the main design issues for RF-EHNs.
The survey is organized as following. The next section presents an overview of RF-EHNs with the focus on the system architecture, RF energy sources and harvesting techniques as well as existing applications. 
Section III introduces the principle and hardware implementation of RF harvesting devices. Then, we introduce the main design issues with single-hop RF-EHNs, multi-antenna RF-EHNs, and multi-hop RF-EHNs in Section IV, Section V, and Section VI, respectively.  We also shed light on the arising challenges in RF-powered cognitive radio networks in Section VII. Then, in Section VIII, we review existing communication protocols designed exclusively for RF-EHNs. Section~IX envisions the future directions of RF-EHNs and discuss the practical challenges. Finally, Section~X concludes this survey.

\section{Overview of RF Energy Harvesting Networks}
\label{sec:overview}

In this section, we first describe the general architecture of an RF-EHN and introduce the RF energy harvesting technique. Then, we review the existing applications of RF-EHNs.

\subsection{Architecture of RF Energy Harvesting Network}

\begin{figure*}
\centering
\includegraphics[width=0.99\textwidth]{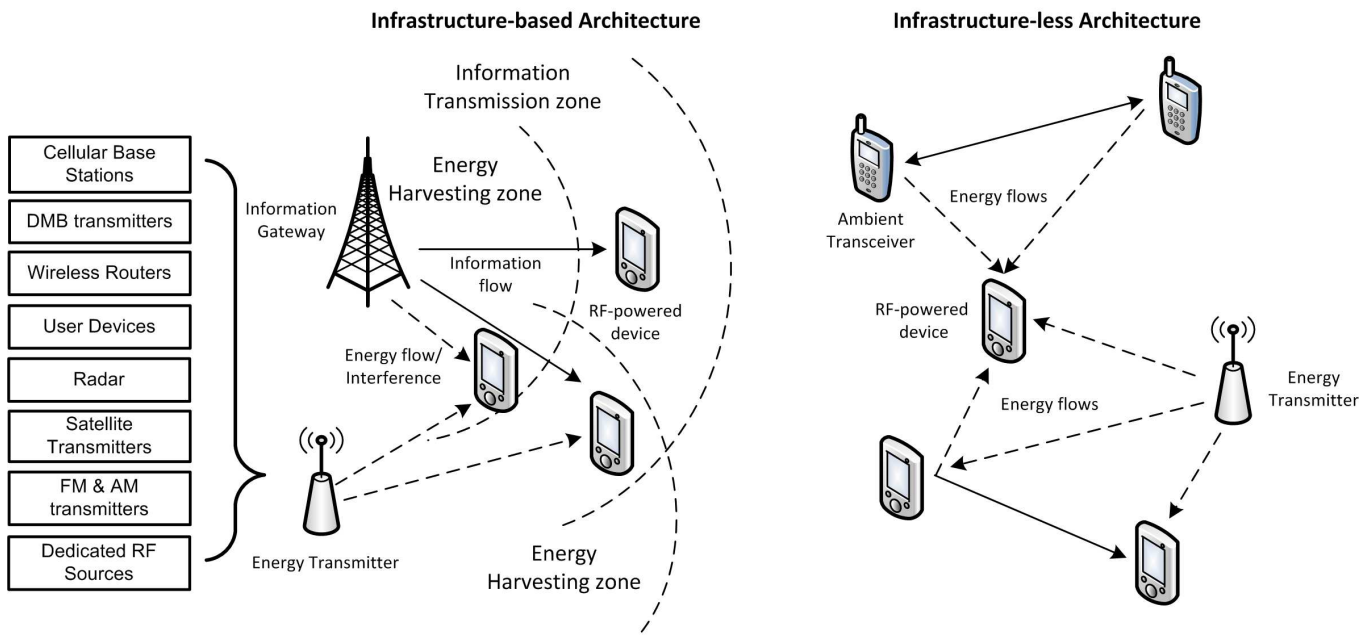}
\caption{A general architecture of an RF energy harvesting network.} \label{network_architecture}
\end{figure*}

A typical centralized architecture of an RF-EHN, as shown in Fig.~\ref{network_architecture}, has three major components, i.e., information gateways, the RF energy sources and the network nodes/devices. The information gateways are generally known as base stations, wireless routers and relays. The RF energy sources can be either dedicated RF energy transmitters or ambient RF sources (e.g., TV towers). The network nodes are the user equipments that communicate with the information gateways. Typically, the information gateways and RF energy sources have continuous and fixed electric supply, while the network nodes harvest energy from RF sources to support their operations. In some cases, the information gateway and RF energy source can be the same. As shown in Fig.~\ref{network_architecture}, the solid arrow lines represent information flows, while the dashed arrow lines mean energy flows. 

The information gateway has an energy harvesting zone and an information transmission zone represented by the dashed circles in Fig.~\ref{network_architecture}. The devices in the energy harvesting zone are able to harvest RF energy from the information gateway. The devices in the information transmission zone can successfully decode information transmitted from the gateway. Generally, the operating power of the energy harvesting component is much higher than that of the information decoding component. Therefore, the energy harvesting zone is smaller than the information transmission zone. Note that the decentralized RF-EHN also has a similar architecture to that shown in Fig.~\ref{network_architecture} except that the network nodes communicate among each other directly. 

\begin{figure*}
\centering
\includegraphics[width=0.95\textwidth]{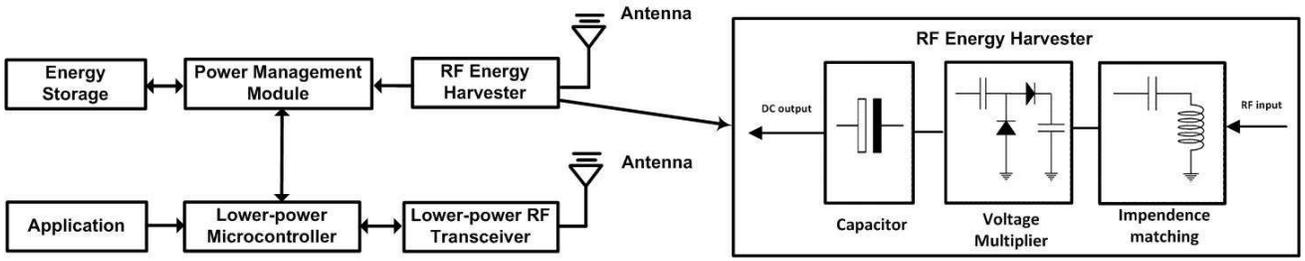}
\caption{A general architecture of an RF energy harvesting device.} \label{node_architecture}
\end{figure*}

Figure~\ref{node_architecture} also shows the block diagram of a network node with RF energy harvesting capability. An RF energy harvesting node consists of the following major components:
\begin{itemize}
\item The application, 
\item A low-power microcontroller, to process data from the application, 
\item A low-power RF transceiver, for information transmission or reception, 
\item An energy harvester, composed of an RF antenna, an impedance matching, a voltage multiplier and a capacitor, to collect RF signals and convert them into electricity, 
\item A power management module, which decides whether to store the electricity obtained from the RF energy harvester or to use it for information transmission immediately, and 
\item An energy storage or battery. 
\end{itemize}

The power management module can adopt two methods to control the incoming energy flow, i.e., \emph{harvest-use} and \emph{harvest-store-use}. In the \emph{harvest-use} method, the harvested energy is immediately used to power the network node. Therefore, for the network node to operate normally, the converted electricity has to constantly exceed the minimum energy demand of the network node. Otherwise, the node will be disabled. In the \emph{harvest-store-use} method, the network node is equipped with an energy storage or a rechargeable battery that stores the converted electricity. Whenever the harvested energy is more than that of the node's consumption, the excess energy will be stored in the battery for future use.
 
Figure~\ref{node_architecture} illustrates the block diagram of an RF energy harvester. 
\begin{itemize}
	\item The antenna can be designed to work on either single frequency or multiple frequency bands, in which the network node can harvest from a single or multiple sources simultaneously. Nevertheless, the RF energy harvester typically operates over a range of frequencies since energy density of RF signals is diverse in frequency. 
	\item The impedance matching is a resonator circuit operating at the designed frequency to maximize the power transfer between the antenna and the multiplier. The efficiency of the impedance matching is high at the designed frequency. 
	\item The main component of the voltage multiplier is diodes of the rectifying circuit which converts RF signals (AC signals in nature) into DC voltage. Generally, higher conversion efficiency can be achieved by diodes with lower built-in voltage. The capacitor ensures to deliver power smoothly to the load. Additionally, when RF energy is unavailable, the capacitor can also serve as a reserve for a short duration. 
\end{itemize}
The efficiency of the RF energy harvester depends on the efficiency of the antenna, the accuracy of the impedance matching between the antenna and the voltage multiplier, and the power efficiency of the voltage multiplier that converts the received RF signals to DC voltage.

For the general node architecture introduced above, the network node has the separate RF energy harvester and RF transceiver. Therefore, the node can perform energy harvesting and data communication simultaneously. In other words, this architecture supports both \emph{in-band} and \emph{out-of-band} RF energy harvesting. In the in-band RF energy harvesting, the network node can harvest RF energy from the same frequency band as that of data communication. By contrast, in the out-of-hand RF energy harvesting, the network node harvests RF energy from the different frequency band from that used for data communication. Since RF signals can carry energy as well as information, theoretically RF energy harvesting and information reception can be performed from the same RF signal input. This is referred to as the simultaneous wireless information and power transfer (SWIPT)~\cite{ZhangRuiMIMO} concept. This concept allows the information receiver and RF energy harvester to share the same antenna or antenna array. 

\subsection{RF Energy Propagation Models}

In RF energy harvesting, the amount of energy that can be harvested depends on the transmit power, wavelength of the RF signals and the distance between an RF energy source and the harvesting node. The harvested RF power from a transmitter in free space can be calculated based on the Friis equation~\cite{balanis2012antenna} as follows:
\begin{eqnarray}
P_{R}=P_{T} \frac{G_{T}G_{R}\lambda^{2}}{(4\pi d)^{2}L}	,
\label{Friis}
\end{eqnarray}
where $P_{R}$ is the received power, $P_{T}$ is the transmit power, $L$ is the path loss factor, $G_{T}$ is the transmit antenna gain, $G_{R}$ is the receive antenna gain, $\lambda$ is the wavelength emitted, and $d$ is the distance between the transmit antenna and the receiver antenna.  

The free-space model has the assumption that there is only one single path between a transmitter and a receiver. 
However, due to RF scattering and reflection, a receiver may collect RF signals from a transmitter from multiple paths.
The two ray ground model captures this phenomenon by considering the received RF signals pass through a line-of-sight path and a reflected path separately. The harvested RF power from a transmitter according to the two ray ground model is given by
\begin{eqnarray}
P_{R}=P_{T} \frac{G_{T}G_{R}h^2_t h^2_r}{ d^{4}L}	,
\label{two-ray}
\end{eqnarray}
where $h_t$ and $h_r$ are the heights of the transmit and receive antennas, respectively. 

The above two deterministic models characterize RF propagation based on determinate parameters. By contrast, probabilistic models draw parameters from a distribution, while allows a more realistic modeling. A practical and widely adopted probabilistic model is a Rayleigh model \cite{S.2001Rappaport}, which represents the situation when there is no line-of-sight channel between a transmitter and receiver. In the Rayleigh model, we have
\begin{eqnarray}
P_{R}= P^{det}_{R} \times 10^{L}\times log(1-unif(0,1))	,
\end{eqnarray}
where  $P^{det}_{R}$ represents the received RF power calculated by a deterministic model. The path loss factor $L$ is defined as $L=-\alpha \log10(d/d_0)$, where $d_0$ is a reference distance. $unif(0,1)$ denotes a random number generated following uniform distribution between 0 and 1. 

The above has presented three common RF propagation models. The aggregated harvested RF energy can be calculated based on the adoption of the network model and RF propagation model.  
Readers can refer to \cite{K2003Sarkar} for more detailed survey of RF propagation models in different environments.

\subsection{RF Energy Harvesting Technique}

Unlike energy harvesting from other sources, such as solar, wind and vibrations, RF energy harvesting has the following characteristics:

\begin{itemize}

\item RF sources can provide controllable and constant energy transfer over distance for RF energy harvesters. 

\item In a fixed RF-EHN, the harvested energy is predictable and relatively stable over time due to fixed distance.

\item Since the amount of harvested RF energy depends on the distance from the RF source, the network nodes in the different locations can have significant difference in harvested RF energy. 

\end{itemize}

The RF sources can mainly be classified into two types, i.e., dedicated RF sources and ambient RF sources.

\subsubsection{Dedicated RF sources} 

Dedicated RF sources can be deployed to provide energy to network nodes when more predictable energy supply is needed. The dedicated RF sources can use the license-free ISM frequency bands for RF energy transfer. The Powercaster transmitter \cite{Powercast} operating on 915MHz with 1W or 3W transmit power is an example of a dedicated RF source, which has been commercialized. However, deploying the dedicated RF sources can incur high cost for the network. Moreover, the output power of RF sources must be limited by regulations, such as Federal Communications Commission (FCC) due to safety and health concern of RF radiations. For example, in the $900MHz$ band, the maximum threshold is 4W \cite{FCC}. Even at this highest setting, the received power at a moderate distance of 20m is attenuated down to only 10 $\mu$W. Due to this limitation, many dedicated RF sources may need to be deployed to meet the user demand. As the RF energy harvesting process with dedicated RF sources is fully controllable, it is more suitable to support applications with QoS constraints. Note that the dedicated RF sources could be mobile, which can periodically move and transfer RF energy to network nodes. In \cite{Erol-Kantarci2012Suresense,Erol-Kantarci2012, Erol-Kantarci2012DRIFT}, different RF energy transmission schemes for mobile power transmitters to replenish wireless sensor networks are investigated.

\subsubsection{Ambient RF sources} 

Ambient RF sources refer to the RF transmitters that are not intended for RF energy transfer. This RF energy is essentially free. 
The transmit power of ambient RF sources varies significantly, from around $10^{6}$W for TV tower, to about 10W for cellular and RFID systems, to roughly 0.1W for mobile communication devices and WiFi systems. 
Ambient RF sources can be further classified into static and dynamic ambient RF sources. 

\begin{itemize}

\item {\em Static ambient RF sources:} Static ambient RF sources are the transmitters which release relatively stable power over time, such as TV and radio towers. Although the static ambient RF sources can provide predictable RF energy, there could be long-term and short-term fluctuations due to service schedule (e.g., TV and radio) and fading, respectively. Normally, the power density of ambient RF sources at different frequency bands is small. As a result, a high gain antenna for all frequency bands is required. Moreover, the rectifier must also be designed for wideband spectrum. In \cite{I1404.4822Flint}, the performance analysis of a sensor powered by static ambient RF sources is performed using a stochastic geometry approach. An interesting finding is that when the distribution of ambient RF sources exhibits stronger repulsion, larger RF energy harvesting rate can be achieved at the sensor.

\item {\em Dynamic ambient RF sources:} Dynamic ambient RF sources are the RF transmitters that work periodically or use time-varying transmit power (e.g., a WiFi access point and licensed users in a cognitive radio network). The RF energy harvesting from the dynamic ambient RF sources has to be adaptive and possibly intelligent to search for energy harvesting opportunities in a certain frequency range. The study in~\cite{SLee2013} is an example of energy harvesting from dynamic ambient RF sources in a cognitive radio network. A secondary user can harvest RF energy from nearby transmitting primary users, and  can transmit data when it is sufficiently far from primary users or when the nearby primary users are idle. 
\end{itemize}

Table~\ref{Data} shows from experiment the amount of harvested RF energy from different sources. We can clearly observe that the energy harvesting rate varies significantly depending on the source power and distance. Typically, the amount of harvested energy is in order of micro-watts, which is sufficient for powering small devices.


\begin{table*}\footnotesize
\centering
\caption{\footnotesize Experimental data of RF Energy Harvesting.} \label{Data}
\begin{tabular}{|l|l|l|l|l|}
\hline
\footnotesize {\bf Source} & {\bf Source Power} & {\bf Frequency} & {\bf Distance} & {\bf Energy Harvested Rate} \\ \hline
\hline
Isotropic RF transmitter \cite{T2008Le} & 4W & 902-928MHz & 15m & 5.5$\mu$W \\
\hline
Isotropic RF transmitter \cite{M2013Stoopman}  & 1.78W & 868MHz & 25m & 2.3$\mu$W \\
\hline
Isotropic RF transmitter \cite{M2014Stoopman} &  1.78W & 868MHz & 27m & 2$\mu$W \\
\hline
TX91501 Powercaster transmitter~\cite{Murtala2012} & 3W & $915MHZ$ & $5m$ & 189$\mu$W  \\
\hline
TX91501 Powercaster transmitter~\cite{Murtala2012} & 3W & 915MHz & 11m & 1$\mu$W \\
\hline
KING-TV tower \cite{Sample2013} & 960kW & 674-680MHz & 4.1km &  60$\mu$W \\
\hline
\end{tabular}
\end{table*}

\subsection{Existing Applications of RF Energy Harvesting}

Wireless sensor networks have become one of the most widely applied applications of RF-EHNs. An RF energy harvester can be used in a sensor node to supply energy. For example, in \cite{G2014Papotto}, the authors design an RF-powered transmitter that supports 915MHz downlink and 2.45GHz uplink bands. An average data rate of 5kbps is achieved, while the maximum instant data rate is up to 5Mbps. The transmitter can be operated with an input power threshold of -17.1 dBm and a maximum transmit power of -12.5 dBm. Various prototype implementations of sensor nodes powered by RF energy are also presented in \cite{Nishimoto2010,Popovic2013,D2012Dondi,Pavone2012,F2012Zhang,A2013Al-Khayari,M2012Al-Lawati,T2013BLim,Farinholt2009}. In \cite{K2013Kaushik,P2012Olds,Seah2013}, a multi-hop RF-powered wireless sensor network is demonstrated through experiments. 
 
The RF-powered devices also have attractive healthcare and medical applications such as wireless body network. Benefiting from RF energy harvesting, low-power medical devices can achieve real-time work-on-demand power from dedicated RF sources, which further enables a battery-free circuit with reduced size. In~\cite{Zhang2010}, the authors design the RF-powered energy-efficient application-specific integrated circuit, featured with a work-on-demand protocol. In \cite{N2013Barroca}, the authors present a body device circuit dual-band  operating at GSM 900 and GSM 1800. The antenna achieves gains of the order 1.8-2.06 dBi and efficiency of $77.6-84\%$. Similar implementations of body devices can also be found in \cite{F2012Zhang,L2014Xia,V2013Kuhn,Y2013Zhang,S2009Mandal}.

Another RF energy harvesting application that has caught intensive research investigation is RFID, widely used for identification, tracking, and inventory management \cite{Y2010Zuo}. Recent developments in low-power circuit and RF energy harvesting technology can extend the lifetime and operation range of conventional RFID tags. In particular, RFID tags, instead of relying on the readers to activate their circuits passively, can harvest RF energy and perform communication  actively.
Consequently, RFID technology has evolved from simple passive tags to smart tags with newly introduced features such as sensing and on-tag data processing and intelligent power management \cite{U2010Olgun}. 
Research progress has covered the designs of RFID tags with RF energy harvesting in rectenna \cite{U2010Olgun,U2010AugOlgun,SSB2012Hong}, rectifier \cite{P2013Kamalinejad,MR2012Shokrani}, RF-to-DC converter \cite{S2012Scorcioni,C2010Zhang}, charge
pump \cite{S2008Amini,S2012Shabana,D2013DDonno} and power harvester \cite{J2009Wilas,A2010Costanzo,A2013Chasin}.

Other than the above popular applications, devices powered by ambient RF energy is attracting increasingly research attention. 
For example, reference \cite{V2013Liu} demonstrates that an information rate of 1 kbps can be achieved between two prototype devices powered by ambient RF signals, at the distance of up to 2.5 feet and 1.5 feet for outdoors and indoors, respectively.
Existing literature has also presented many implementations of battery-free devices powered by ambient energy from WiFi \cite{U2012Olgun1,U2012Olgun}, GSM  \cite{M2011Arrawatia,M2013Pinuela,W2012M,Batool2012} and DTV bands \cite{P2013Nintanavongsa,S2012Keyrouz, C.sept.Mikeka,H2010Nishimoto,R2012Vyas,S2012Kitazawa}
as well as ambient mobile electronic devices \cite{B2011G}.

Additionally, RF energy harvesting can be used to provide charging capability for a wide variety of low-power mobile devices such as electronic watches, hearing aids, and MP3 players, wireless keyboard and mouse, as most of them consume only micro-watts to milli-watts range of power. In \cite{Jabbar2010}, the authors present a design of an RF circuit that enables continuous charging of mobile devices especially in urban areas where the density of ambient RF sources is high. 

\section{Circuit Design}

This section introduces some background related to the hardware circuit designs of RF energy harvesting devices. 
Here, the purpose is to introduce some basic kownledge of circuit design required to understand the communication aspects of the RF-EHN. Again, the comprehensive survey of the works related to circuit design and electronics for RF energy harvesting is beyond the scope of this paper.



\begin{table*}\footnotesize
\centering
\caption{\footnotesize Circuit Performance Comparison.} \label{circuit_performance}
\begin{tabular}{|l|p{3.1cm}|p{3.5cm}|l|l|}
\hline
\footnotesize {\bf Literature} & {\bf Minimum RF Input Power $@$ Output Voltage}  & {\bf Peak Conversion Efficiency $@$ RF Input Power} & {\bf Frequency} & {\bf Technology} \\ \hline
\hline
F. Kocer, {\em et al} \cite{F2006Kocer} (2006)   & -19.58 dBm @ 1V & 10.9$\%$ @ -12 dBm & 450MHz & 0.25$\mu$ m CMOS  \\
\hline
J. Yi, {\em et al} \cite{J2007Yi} (2006) & N. A. & 26.5$\%$ @ -11 dBm & 900 MHz  & 0. 18$\mu$ m CMOS   \\
\hline
S. Mandal, {\em et al} \cite{S2007Mandal} (2007) & -17.7 dBm @ 
0.8V & 37$\%$ @ -18.7 dBm & 970 MHz & 0. 18$\mu$m CMOS  \\
\hline
A. Shameli, {\em et al} \cite{A2007Shameli} (2007) & -14.1 dBm @ 1V & N. A. & 920MHz & 0.18$\mu$ m CMOS \\
\hline
T. Le, {\em et al} \cite{T2008Le} (2008) & -22.6 dBm @ 
1V & 30$\%$ @ -8 dBm  & 906MHz & $0.25\mu$ m CMOS  \\
\hline
Y. Yao, {\em et al} \cite{Y2009Yao} (2009) & -14.7 dBm @ 1.5V & $15.76\%$ @ 12.7 dBm  & 900MHz & $0.35\mu$ m CMOS  \\
\hline
T. Salter, {\em et al} \cite{T2009Salter} (2009) & -25.5 dBm @ 1V & N. A. & 2.2GHz & 130nm CMOS  \\
\hline
G. A. Vera, {\em et al} \cite{G2010AndiaVera} (2010) & N. A. & $42.1\%$ @ -10 dBm  & 2.45GHz & SMS 7630 \\
\hline
G. Papotto, {\em et al} \cite{G2011Papotto} (2011) & -24 dBm ($4\mu$ W) @ 1V  & 11$\%$  @ -15 dBm  & 915MHz  &  90nm CMOS   \\
\hline
O. H. Seunghyun, {\em et al} \cite{O2012h} (2012) & -32 dBm @ 1V & N. A. & 915MHz & 130nm CMOS  \\
\hline
J. Masuch, {\em et al} \cite{J2012Masuch} (2012) & N. A. & 22.7$\%$ @ -3 dBm   & 2.4GHz  & 130nm CMOS   \\
\hline
S. Scorcioni, {\em et al} \cite{S2012Scorcioni868} (2012) & -17 dBm @ 2V& 60$\%$ @ -3 dBm & 868MHz & 0.130$\mu$ m CMOS  \\
\hline
S. Scorcioni, {\em et al} \cite{S2012AScorcioni} (2012) & -17 dBm @ 2V  & 60$\% $ 55$\%$ @ -10 dBm & 868MHz & 0.13$\mu$ m CMOS   \\
\hline
T. Taris, {\em et al} \cite{T2012Taris} (2012) & -22.5 dBm @ 0.2V & N. A. & 900MHz &  HSMS-2852 \\
 &  -11 dBm @ 1.08V & & &  \\
\hline
H. Sun, {\em et al} \cite{H2012Sun} (2012) & -3.2 dBm @ 1V & 83$\%$ @ -1 dBm  & 2.45GHz & HSMS-2852 \\
\hline
 
D. Karolak, {\em et al} \cite{D2012Karolak} (2012) & -21 dBm @ 1.45V  & 65.2 $\%$ @ -21 dBm & 900MHz & 13nm CMOS \\
&  -21 dBm @ 1.43V  & 64 $\%$ @ -21 dBm & 2.4GHz &  \\
\hline
M. Roberg {\em et al} \cite{M2012Roberg} (2012) & 40 dBm @ 30V & 85$\%$ @ 40 dBm & 2.45GHz & SMS-7630 \\
\hline

P. Nintanavongsa {\em et al} \cite{P2012Nintanavongsa} (2012) & -10 dBm @ 1V &  10$\%$ @ -10 dBm  & 915MHz & SMS-2852 \\
\hline
Bruno R. Franciscatto, {\em et al} \cite{B2013R} (2013) &  0 dBm @ 1.2V  & $70.4\%$ @ 0 dBm & 2.45GHz & HSMS-2855  \\
\hline
S. Scorcioni, {\em et al} \cite{S2013Scorcioni} (2013) &  -16 dBm @ 2V  & $ 58\%$ @ -3 dBm & 868MHz & 130nm CMOS  \\
\hline
M. Stoopman, {\em et al} \cite{M2013Stoopman} (2013) & -26.3 dBm @
1V & $31.5\%$ @ -15 dBm & 868MHz & 90nm CMOS  \\
\hline
X. Wang, {\em et al} \cite{W2013Xiaoyu} (2013) & -39 dBm @ 2.5V  & N. A. & AM frequency
band & N. A.  \\
\hline
T. Thierry, {\em et al} \cite{T2013Thierry} (2013) & -10 dBm @ 2.2V & N. A.& 900MHz & HSHS-2852  \\
& - 20 dBm @ 0.4V  & & 2.4GHz &  \\
\hline
A. Nimo, {\em et al} \cite{A2013Nimo} (2013) &  -30 dBm @ 1.9 V   & $55\%$ @ -30 dBm  & 13.56MHz & HSMS-286B  \\
\hline
S. B. Alam, {\em et al} \cite{S2013BinAlam} (2013) & -15 dBm @ 0.55V & N. A. & 2.45GHz & HSMS-2850 \\
\hline
S. Agrawal, {\em et al} \cite{S2014Agrawal} (2014) & -10 dBm @ 1.3V & $ 75\%$ @ -10 dBm  & 900MHz  & HSMS-2852 \\
\hline
M. Stoopman, {\em et al} \cite{M2014Stoopman} (2014) & - 27 dBm @ 1V & $40\% $@ -17 dBm &  868MHz  & 90nm CMOS  \\
 \hline
\end{tabular}
\end{table*}

\subsection{Circuitry Implementations}

There have been a large number RF energy harvester implementations based on various different technologies such as CMOS, HSMS and SMS. Table~\ref{circuit_performance} shows the circuit performance of some up-to-date designs. Most of the implementations are based on the CMOS technology. Generally, to achieve 1V DC output, -22 dBm to -14 dBm harvested RF power is required. Though CMOS technology allows a lower minimum RF input power, the peak RF-to-DC conversion efficiency is usually inferior to that of HSMS technology. The efficiency above $70\%$ can be achieved when the harvested power is above -10 dBm.  
For RF energy harvesting at a relatively high power (e.g., 40 dBm/10W), SMS technology can be adopted. In particular, as shown in \cite{M2012Roberg}, 30V output voltage is achieved at 40 dBm input RF power with $85\%$ conversion efficiency. However, when the harvested RF power is low, the conversion efficiency is low. For example, only $10\%$ as input power is -10 dBm \cite{P2012Nintanavongsa}.

\subsection{Antenna Design}

An antenna is responsible for capturing RF signals. Miniaturised size and high antenna gain are the main aims of antenna technology.  The authors in \cite{Y2013Choi} report a comparative study of  several antenna designs for RF energy harvesting. Several antenna topologies for RF energy harvesting have been reported in \cite{Thomas2006}. In \cite{A2013Aziz}, the authors perform a comparison of existing antenna structures. Antenna array design has also been studied for effective RF energy harvesting in \cite{X2011Shao,JM2012Barcak}. Antenna arrays are effective in increasing the capability for low input power. However, a tradeoff exists between antenna size and performance.

For hardware implementations, research efforts have been made for narrow-band antenna (typically from several to tens of MHz) designs in a single band \cite{M2011Arrawatia,S2013BinAlam,M2013Arsalan,M2010Arrawatia}, and dual bands \cite{Z2013SeptZakaria,B2013Li,X2011Shao,Z2013Zakaria} as well as triple bands \cite{B2013L,D2011Masotti,Keyrouz2013}. Moreover, broadband antenna designs (typical on order of 1GHz) have been the focus of some recent work    \cite{D2013Yi,A2011Buonanno,A2012Nimo,D2012Yi,J2012Zhang,J2013Zhang,N2013A,L2012Vincetti}.

\subsection{Matching Network}

The crucial task of matching network  is to reduce the transmission loss from an antenna to a rectifier circuit and increase the input voltage of a rectifier circuit \cite{Agrawal2013}. To this end, a matching network is usually made with reactive components such as coils and capacitors that are not dissipative \cite{M2011T}. Maximum power transfer can be realized  when the impedance at the antenna output and the impedance of the load are conjugates of each other. This procedure is known as \emph{impedance matching}. Currently, there exist three main matching network circuits designed for RF energy harvesting, i.e., \emph{transformer, shunt inductor, LC network}. The detailed introduction of these circuits can be found in \cite{M2011T}.

\subsection{Rectifier}

The function of a rectifier is to convert the input RF signals (AC type) captured by an antenna into DC voltage. A major challenge of the rectifier design is to generate a battery-like voltage from very low input RF power. Generally, there are three main options for a rectifier, which are a diode \cite{J2004A}, a bridge of diodes \cite{M2004Ghovanloo} and a voltage rectifier multiplier \cite{J2007P}.  

The diode is the main component of a rectifier circuit. The rectification performance of a rectifier mainly depends on the saturation current, junction capacitance and its conduction resistance of the diode(s) \cite{M2011T}. The circuit of a rectifier, especially the diode, determines the RF-to-DC conversion efficiency. The most commonly used diode for rectennas is silicon Schottky barrier diodes. Generally, a diode with a lower built-in voltage can achieve a higher rectifying efficiency. This is because larger voltage will result in significantly more harmonic signals due to the nonlinear characteristics of the diode, thus notably decreasing the rectifying efficiency \cite{J2005A}.
In \cite{Y2013Wu}, a model is developed to characterize the RF-to-DC rectification with low input power.  
Based on the model, the authors derive closed-form solutions for the equilibrium voltage and the input resistance of the rectifier. A quasi-static model is further developed to describe the dynamic charging of the capacitor of the rectifier.

\subsection{Receiver Architecture Design}

The traditional information receiver architecture designed for information reception may not be optimal for SWIPT. The reason is because information reception and RF energy harvesting works on very different power sensitivity (e.g., -10 dBm for energy harvesters versus -60 dBm for information receivers) \cite{ZhangRuiMIMO}.  
This inspires the research efforts in devising the receivers for RF-power information receivers. Currently, there are four typical types of receiver architectures.

\begin{figure} 
    \centering
    \subfigure [Separated Receiver Architecture] {
     \label{SRA}
     \centering
     \includegraphics[width=0.38 \textwidth]{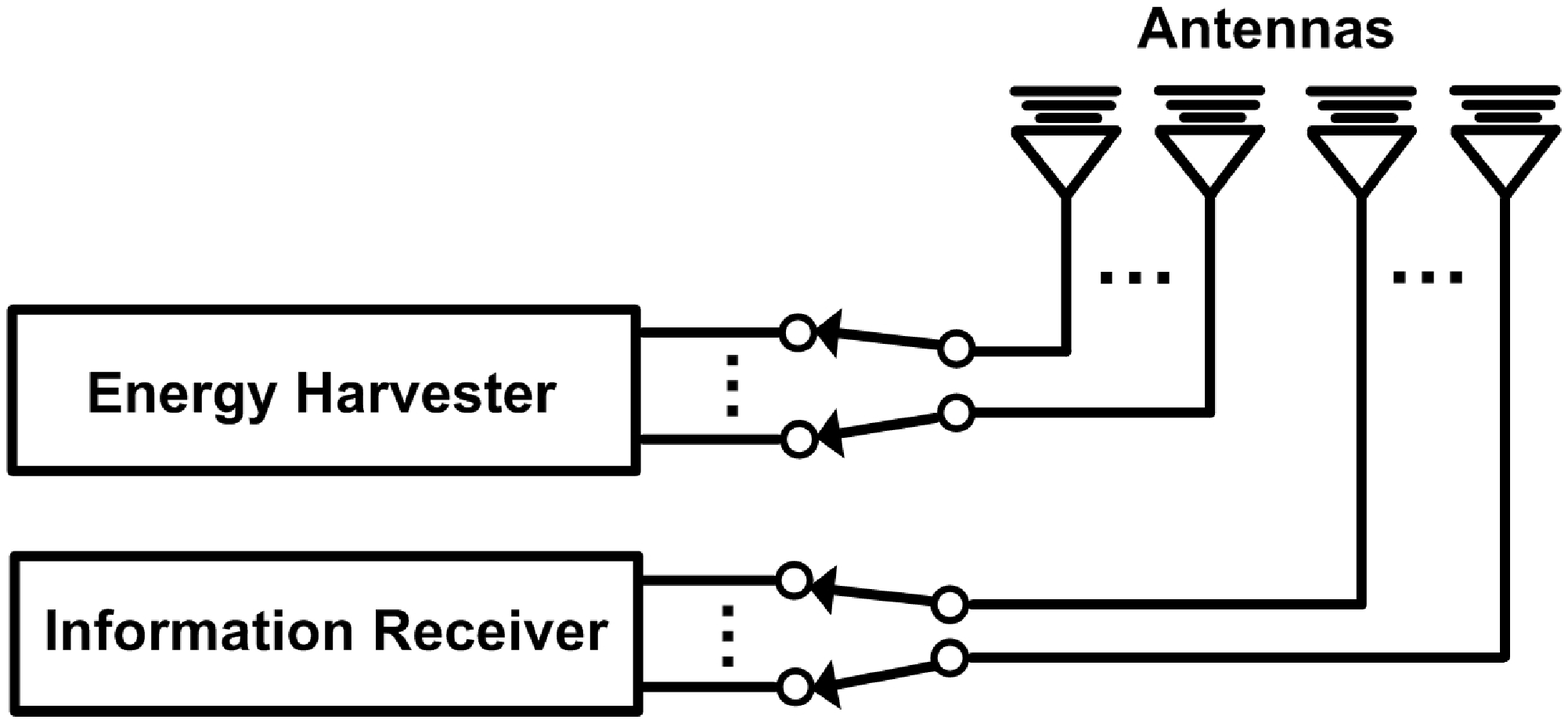}} \\
 \centering
  \subfigure [Time Switching Achitecture] {
   \label{TSA}
   \centering
   \includegraphics[width=0.35 \textwidth]{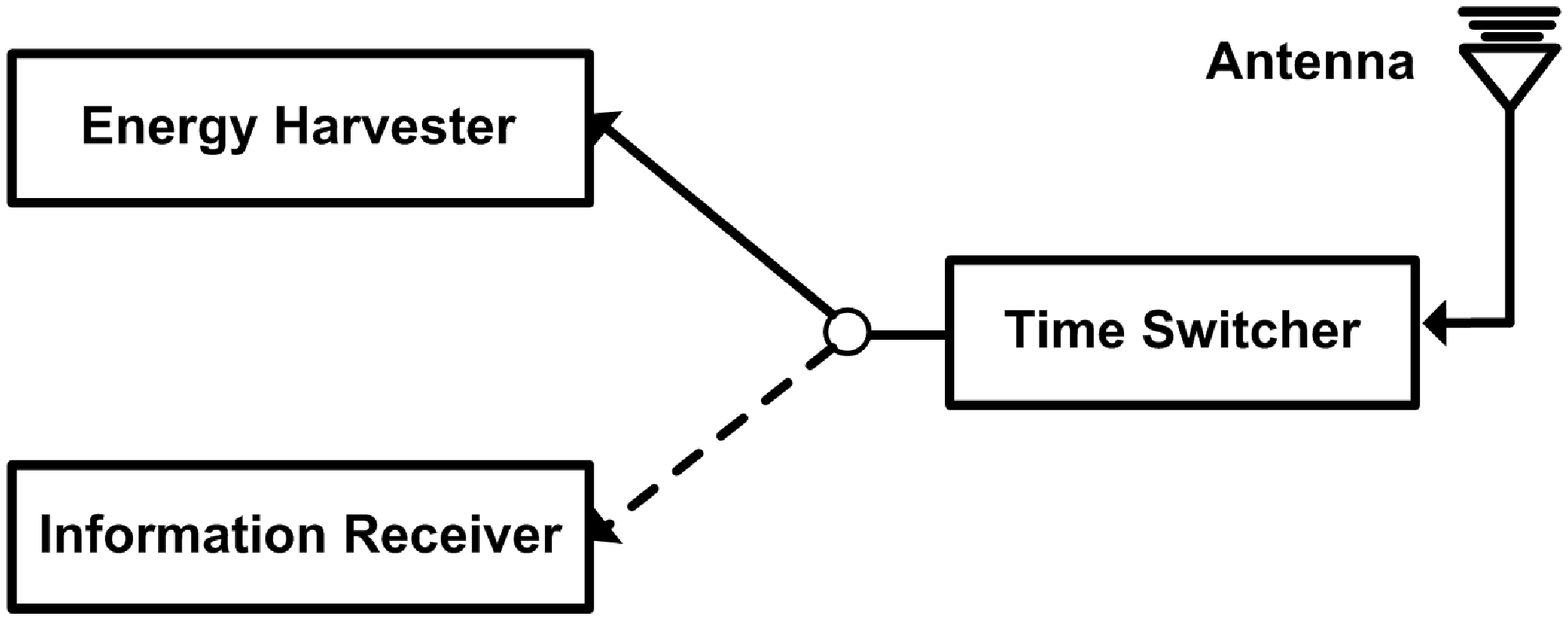}} \\
  \centering
 \subfigure [Power Splitting Achitecture] {
  \label{PSA}
  \centering
  \includegraphics[width=0.35 \textwidth]{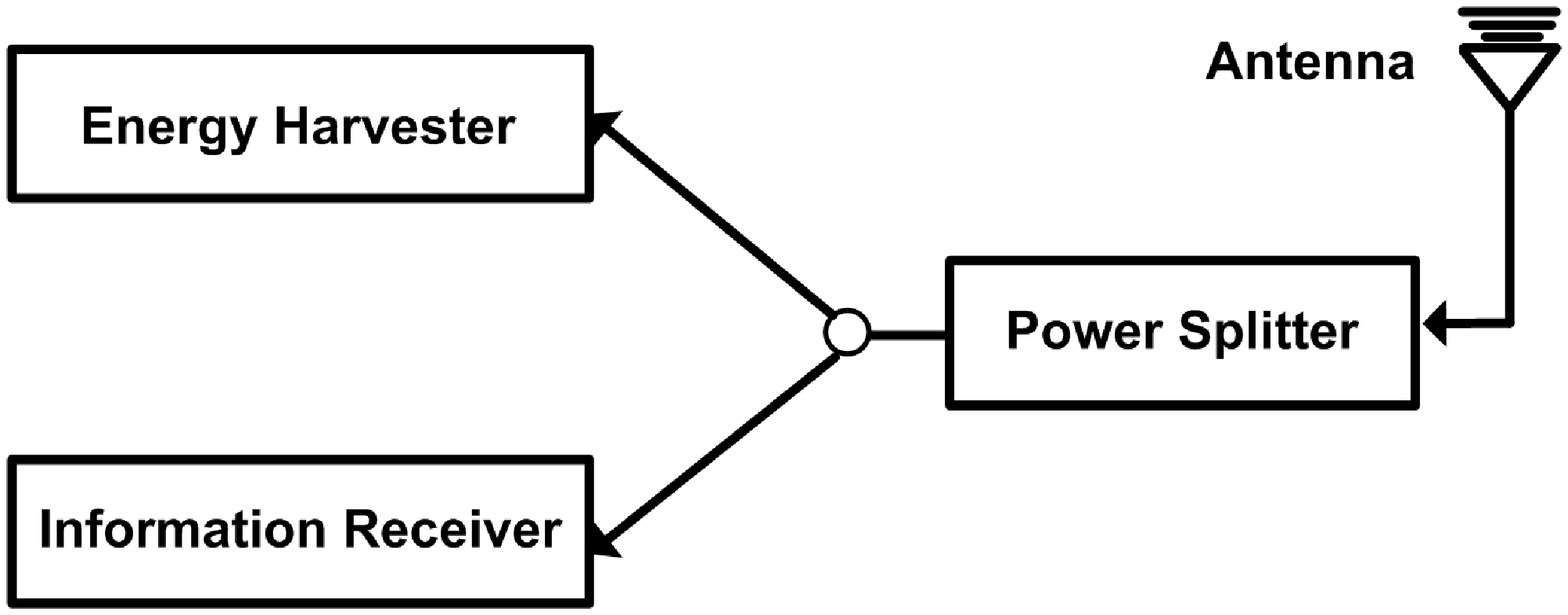}}\\
 \centering
 \subfigure [Ingerated Receiver Achitecture] {
  \label{IRA}
  \centering
  \includegraphics[width=0.45 \textwidth]{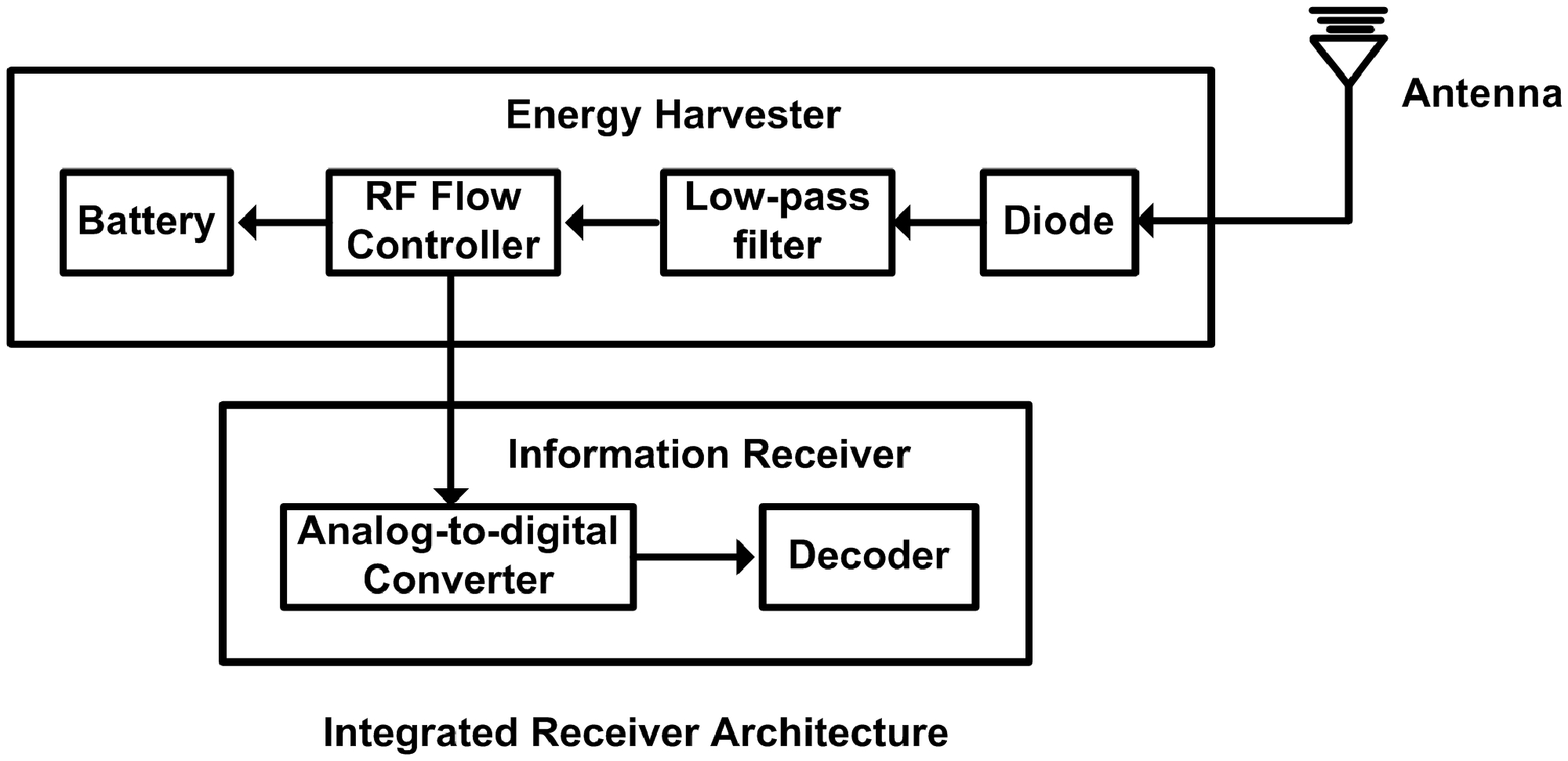}}
 \caption{Receiver Architecture Designs for RF-powered Information Receiver}
 \label{receiver_designs}
 \end{figure}

\begin{itemize}

\item Separated Receiver Architecture: Separated receiver architecture, also known as antenna-switching \cite{ZhangRuiMIMO}, equips an energy harvester and information receiver with independent antenna(s) so that they observe different channels. Figure~\ref{SRA} shows the model for the separated receiver architecture.
The antenna array is divided into two sets with each connected to the energy harvester or the information receiver. Consequently, the architecture allows to perform energy harvesting and information decoding independently and concurrently. 
The antenna-switching scheme \cite{ZhangRuiMIMO} can be used to optimize the performance of the separated receiver architecture. 

\item Co-located Receiver Architecture: The co-located receiver architecture let an energy harvester and an information receiver share the same antenna(s) so that they observe the same channel(s). As a single antenna can be adopted, the co-located receiver architecture is able to enable a smaller size compared to the separated receiver architecture.
This architecture can be categorized into two models, i.e., time-switching and power-splitting architectures. The time-switching architecture, as shown in Fig.~\ref{TSA}, allows the network node to switch and use either the information receiver or the RF energy harvester for the received RF signals at a time.
When a time-switching receiver $j$ working in the energy harvesting mode, the power harvested from source $i$ can be calculated as follows:
\begin{eqnarray}
P_{j,i}=\eta P_{i}|h_{i,j}|^{2}
\end{eqnarray}
where $\eta$ denotes the energy harvesting efficiency factor, $P_{i}$ is the transmit power at source $i$, and $h_{i,j}$ denotes the channel gain between between source $i$ and receiver $j$. Let $W$ and $\sigma^2$ denote the transmission bandwidth and noise power, respectively.
When the time-switching receiver $j$ working in the information decoding mode, the maximum information decoding rate from source $i$ is
\begin{eqnarray}
R_{j,i}=W \log(1+ P_{i}|h_{i,j}|^{2}/\sigma^2).
\end{eqnarray}
In the power-splitting architecture, as shown in Fig.~\ref{PSA}, the received RF signals are split into two streams for the information receiver and RF energy harvester with different power levels.
Let $\theta_{j} \in [0,1]$ denote the power-splitting coefficient for receiver $j$, i.e., $\theta_{j}$ is the fraction of RF signals used for energy harvesting. 
Similarly, the power of harvested RF energy at a power-splitting receiver $j$ from source $i$ can be calculated as follows:
\begin{eqnarray}
P_{j,i}=\eta P_{i}|h_{i,j}|^{2} \theta_{j}  	.
\end{eqnarray}
Let $\sigma^2_{SP}$ denote the power of signal processing noise. The maximum information decoding rate at the power-splitting receiver $j$ decoded from source $i$ is
\begin{eqnarray}
R_{j,i}=W\log	\Bigg(1+(1-\theta_{i})P_{i}|h_{i,j}|^{2}/(\sigma^2+\sigma^2_{SP})	\Bigg) 	.
\end{eqnarray}
In practice, \emph{power splitting} is based on a power splitter and \emph{time switching} requires a simpler switcher. 
It has been recognized that theoretically power-splitting achieves better tradeoffs between information rate and amount of RF energy transferred \cite{ZhangRuiMIMO,XZhou2013}.

\item Integrated Receiver Architecture: In the integrated receiver architecture proposed in \cite{XZhou2013}, the implementation of RF-to-baseband conversion for information decoding, is integrated with the energy harvester via the rectifier. Therefore, this architecture allows a smaller form factor. Figure~\ref{IRA} demonstrates the model for integrated receiver architecture. Note that the RF flow controller can also adopt a switcher or power splitter, like in the co-located receiver architecture. However, the difference is that the switcher and power splitter are adopted in the integrated receiver architecture.

\item Ideal Receiver Architecture: The ideal receiver architecture assumes that the receiver is able to extract the RF energy from the same signals used for information decoding. However,  
as pointed out in \cite{XZhou2013}, this assumption is not realistic in practice. The current circuit designs are not yet able to extract RF energy directly from the decoded information carrier. In other words, any energy carried by received RF signals sent for an information receiver is lost during the information decoding processing. Existing works that consider ideal receiver architecture, such as \cite{Varshney2008,P2010Grover,AMFouladgar2012,P2013Popovski}, generally analyze the theoretical upper bound of receiver performance.

\end{itemize}

The studies in \cite{XZhou2013} show that when the circuit power consumptions are relatively small compared with the received signal power, the integrated receiver architecture outperforms the co-located receiver architecture at high harvested energy region, whereas the co-located receiver architecture is superior at low harvested energy region. When the circuit power consumption is high, the integrated receiver architecture performs better. It is also shown that for a system without minimum harvested energy requirement, the integrated receiver achieves higher information rate than that of the separated receiver at short transmission distances.

\begin{figure}
\centering
\includegraphics[width=0.5\textwidth]{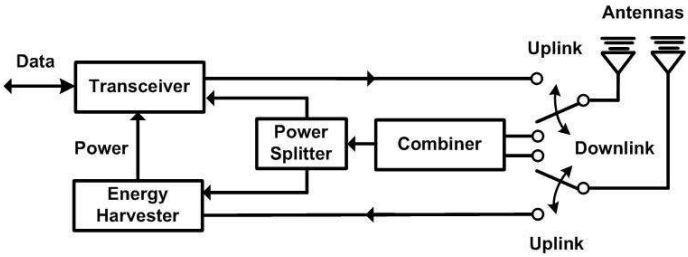}
\caption{An Architecture for Dual-antenna Receiver} \label{dual_antenna}
\end{figure}

With an antenna array, the dual-antenna receiver architecture introduced in \cite{K2013Huang} can be adopted. Illustrated in Fig.~\ref{dual_antenna}, a combiner is adopted to coherently combine the input RF signals for enhancement of the received power. This architecture can be easily extended to the case with a larger number of antennas and the case with time-switching operation.

The following sections will review various communication resource allocation issues and designs.  

\section{Single-hop RF-EHNs}
\subsection{Multi-user Scheduling}

The goals of multi-user scheduling in RF-EHNs are to achieve the best utilization of resources (e.g., RF energy and frequency band) through allocation among different users achieving fairness and to meet some QoS criteria (e.g., throughput, delay or packet loss requirement) under the RF energy harvesting constraint as well as taking into account of implementation complexity and scalability. A major difference from conventional multi-user scheduling is that, RF-EHNs require new criteria in scheduling fairness and energy harvesting requirements. For instance, in conventional wireless communication systems, the maximum normalized SNR scheme, which schedules the user with the maximum instantaneous normalized SNR, can maximize the users' capacities while maintaining proportional fairness among all users \cite{L2006Yang}. However, for the RF-EHN with SWIPT, such a scheduling discipline may fail to fulfill the energy harvesting requirement while guaranteeing fairness, as it leads to the minimum possible harvested energy by the users.  \cite{R1306.3093Morsi}.
The reason is straightforward since the best states of the channels are exploited for information decoding rather than RF energy harvesting.
Therefore, balancing the information and energy tradeoff is one of the primary concerns in designing multi-user scheduling schemes. We categorize the existing scheduling disciplines into three classes: throughput fairness scheduling, throughput maximization scheduling and utilization optimization scheduling.

\subsubsection{Throughput Fairness Scheduling}

The aim of throughput fairness scheduling is to ensure fairness among individual nodes. In \cite{H1304.7886Ju}, a \emph{doubly near-far} phenomenon which results in unfair throughput allocation is found and studied in a multi-user RF-EHN. In particular, the receiver far away from an access point not only harvests less RF energy but also suffers from more signal power attenuation in the uplink information transmission, compared to the nearer receiver. This doubly near-far phenomenon has been considered in different network models \cite{L1312.1450Liu,H1304.7886Ju,H1403.7123Ju}.
In \cite{H1304.7886Ju}, a joint downlink RF energy transfer and uplink information transmission problem is investigated in a multiple access system. The authors propose a \emph{harvest-then-transmit} protocol which allows the receivers to first harvest energy from the downlink broadcast signals, and then use the harvested energy to send independent uplink information to the access point based on time division multiple access (TDMA). To maximize the system throughput, the time allocations for the downlink energy transfer and uplink information transmissions are jointly optimized. To address the doubly near-far problem, the authors define a performance metric called common-throughput. The metric evaluates the constraint that all receivers are assigned with equal throughput regardless of their locations. An iterative algorithm is developed to solve the common-throughput maximization problem. The throughput and fairness tradeoffs for sum-throughput maximization and common-throughput maximization are also analyzed.

The work in \cite{H1403.7123Ju} extends the harvest-then-transmit protocol to the case with user cooperation to overcome the doubly near-far problem. Specifically, considering TDMA-based RF-EHNs with two users, the user located nearer to the access point is allowed to use some of its allocated time and harvested energy to relay the information of the farther user, before transmitting its own information with the rest of time and energy. To maximize the weighted sum-rate of the two users, the time and power allocations for both wireless energy transfer in the downlink and information transmission as well as relaying in the uplink are optimized. Demonstrated by simulations, the proposed protocol with user cooperation increases both the throughput and user fairness considerably.

The study in \cite{L1312.1450Liu} extends  \cite{H1304.7886Ju} by considering the case with a multi-antenna access point. The access point can control the amount of energy transferred to different receivers by adjusting the energy beamforming weights. To cope with the doubly near-far problem, a joint optimization of time allocation, the downlink energy beamforming, uplink transmit power allocation and receive beamforming is designed to maximize the minimum throughput among all receivers. The authors formulate a non-convex problem for an optimal linear minimum-mean-square-error (MMSE) receiver employed at the access point, and solve the problem optimally via a two-stage algorithm. The algorithm first fixes the time allocation and obtains the corresponding optimal downlink energy beamforming and uplink power allocation solution. The algorithm performs the time allocation by solving a sequence of convex feasibility problems and a one-dimension search for energy beamforming and power allocation, respectively. Furthermore, to reduce computational complexity, the authors propose two suboptimal designs based on the zero-forcing receiver in the uplink information transmission. It is found that when the portion of time allocated to RF energy transfer increases or when the distance between users and the access point decreases, the performance of the proposed suboptimal solutions with the zero-forcing receiver approaches that of optimal solution with the MMSE receiver in terms of the max-min throughput. 

The study in \cite{Q1403.3665Sun} considers the same system model as in \cite{H1304.7886Ju} with the \emph{harvest-then-transmit} scheme. A low-complexity fixed point iteration algorithm is proposed for the min-throughput maximization problem. Simulation results show that the fixed point iteration algorithm achieves similar individual throughput compared to the iterative algorithm in \cite{H1304.7886Ju} but with much lower computation complexity.

The authors in \cite{R1401.1943Morsi} study a downlink multi-user scheduling problem for a time-slot based RF-EHN with SWIPT. A protocol which schedules a single user for information reception and the others for energy harvesting in each time-slot is proposed. To control the information and energy tradeoff, the authors propose two scheduling schemes. The first scheme schedules the user according to the descending order of normalized SNR. The second scheme selects the user having the smallest throughput among the set of users whose normalized SNR orders fall into a given predefined set of allowed orders. Both scheduling schemes are shown to achieve proportional fairness in terms of the amount of harvested energy by the users. It is also shown that the lower the selection order for the order-based normalized SNR scheme, or the lower the orders in the predefined set for the order-based equal throughput scheme, the higher the average total amount of harvested energy at the expense of a reduced ergodic system capacity. Furthermore, the authors analyze and give the closed-form expressions for the average per-user harvested energy and ergodic capacity of the proposed schemes for independent and non-identically distributed Rayleigh, Ricean, Nakagami$-m$, and Weibull fading channels.

The authors in \cite{G1403.3991Yang} introduce a frame-based transmission protocol in a massive multi-input multi-output (MIMO) system with imperfect channel state information (CSI). The protocol divides each frame into three phases. The access point estimates downlink channels by exploiting channel reciprocity from the pilots sent by users in uplink channels and broadcasts RF energy to all users in the first and second phases, respectively. Then, the users transmit their independent information with harvested energy to the access point simultaneously in the third phase. The scheme optimizes the time and energy allocation with the aim to maximize the minimum rate among all users. The authors also define a metric called, massive MIMO degree-of-rate-gain, as the asymptotic uplink rate normalized by the logarithm of the number of antennas at the access point. The proposed transmission scheme is shown to be optimal in terms of the metric and is able to guarantee the best possible fairness by asymptotically achieving a common rate for all users.

\subsubsection{Throughput Maximization Scheduling}

The throughput maximization scheduling disciplines are devised to maximize the system throughput under energy harvesting constraint (e.g., circuit-power consumption). Both references \cite{W2013Wang} and \cite{K2013Huang} deal with power allocation problems for system throughput maximization. The authors in \cite{W2013Wang} investigate a MIMO downlink system with co-channel interference. To mitigate the interference, the authors propose to use block diagonalization preceding method which can support a limited number of information receivers due to zero-forcing channel inversion \cite{H2004Spencer}. In this context, only some of the users can be scheduled for information transmission over the same time and frequency block, while other idle users can only be arranged to harvest energy. A power allocation problem is formulated to maximize the throughput of information receivers under the constraints of downlink transmit power and energy harvesting of the idle users. The authors develop a bisection search method for power allocation and define the necessary conditions for existence of optimal solution.

The authors in \cite{K2013Huang} consider an OFDMA broadband system where a multi-antenna base station not only communicates but also transfers RF energy to the users. The throughput maximization problems, under the constraints of circuit-power consumption at the users and transmit power, are formulated for the cases of single user or multiple users. The problems consider fixed or variable information encoding rates, in downlink or uplink. The corresponding power allocation algorithms are designed for the formulated problems.

Both references \cite{X1404.0471Kang} and \cite{H1403.2580Ju} consider multi-user time allocation problems in a full-duplex TDMA-based RF-EHN. The system contains a hybrid access point which is equipped with two antennas and is able to broadcast RF energy in the downlink and receiver information in the uplink simultaneously. In \cite{X1404.0471Kang}, the authors characterize two fundamental optimization problems for the considered system. Specifically, the first problem maximizes the total throughput of the system subject to the time constant. An algorithm with linear complexity is proposed to calculate an optimal time allocation. It is shown that, though constrained by the constant time, the sum of the throughput of the network is non-decreasing with the increase of the number of users. The second problem minimizes the total energy harvesting time and transmission time of the system subject to the data transmission constraints of each user. A two-step algorithm is introduced to obtain an optimal time allocation. It is shown by simulations
that the users with high SNR should have priority to transmit data. The authors in \cite{H1403.2580Ju} aims to maximize the weighted sum-throughput of the considered system, by jointly optimizing the time allocations to the access point and users for downlink energy transfer and uplink information transmission, respectively, as well as the transmit power allocations at the access point.
The authors obtain optimal and suboptimal time and power
allocation solutions for the cases with perfect
and imperfect self-interference cancellation, respectively.
The simulation results show that the full-duplex RF-EHNs is more beneficial than half-duplex ones in the cases when the self-interference can be effectively cancelled, the system contains a sufficiently large number of users, and/or the peak transmit power constraint is more stringent as compared to the average transmit power at the access point.
 
\subsubsection{Utility Optimization Scheduling}
The utility optimization scheduling disciplines are developed to handle various objectives with or without constraints, through centralized or decentralized approaches. Both references \cite{Ng2013OFDM} and \cite{Qi2014Sun} investigate energy efficiency problems. In \cite{Ng2013OFDM}, the authors consider a multi-receiver OFDMA downlink system. A joint design of transmit power allocation and receiver operation based on time-switching is formulated as a mixed non-convex and combinatorial optimization problem. The objective is to maximize energy efficiency (i.e., bit/Joule) of information transmission. The authors introduce an iterative algorithm exploring nonlinear fractional programming and Lagrange dual decomposition to solve the formulated problem. Simulation results demonstrate that, compared with a baseline scheme maximizing the system capacity, the proposed algorithm has better performance in terms of average energy efficiency and system throughput. Besides, the proposed algorithm is shown to achieve higher total harvested energy when increasing the number of receivers, and converge within a small number of iterations. The authors in \cite{Qi2014Sun} aim to maximize the energy efficiency in SWIPT to an information receiver and an energy harvester. Considering statistical CSI feedback, the authors first propose two optimal power allocation algorithms based on gradient
projection and golden section search, respectively. Then, a suboptimal algorithm based on two layer bisection search with low-complexity is devised. The simulation shows that the suboptimal algorithm achieves near-optimal performance. Additionally, it is revealed that SWIPT offers higher energy efficiency compared with conventional information transmission.

The authors in \cite{DNiyatoICC2014} deal with the admission control policy to support QoS in the network. The policy determines the RF energy transfer strategy to maximize the reward of the network, while the throughput of each admitted user is maintained at the target level. An optimization problem based on a Markov decision process is introduced to achieve an optimal admission control policy.

All the scheduling disciplines reviewed above are based on a centralized approach. However, when the size of the system grows, the centralized approaches suffer from the curse of dimensionality, thus causing intractable computation complexity. Although suboptimal solutions (e.g., \cite{L1312.1450Liu}) are introduced to relax complexity, they come along with detrimental performance. Therefore, decentralized approaches are designed to achieve the optimal (e.g., \cite{D2014NiyatoICC} and \cite{T2014Hoang}) or local optimal solution (e.g., \cite{D2014T}) as well as easing the complexity. In \cite{D2014NiyatoICC}, the authors develop a non-cooperative game framework for competition of RF energy in a decentralized wireless network. A bidding strategy based on stochastic response dynamic is proposed for wireless nodes to achieve a Nash equilibrium. 
The authors in \cite{D2014NiyatoCooperation} introduce a coalitional game framework in a delay-tolerant network where RF-powered mobile nodes can cooperatively help one another in packet delivery. Considering that some mobile nodes may secretly make deviation from its coalition, a repeated coalition formation game is developed for the mobile nodes to improve long-term payoff.
In \cite{T2014Hoang}, the authors consider bidding competitions for both radio and energy resources. Specifically, an optimization problem is formulated as a decentralized partially observable Markov decision process with the objective to minimize the total number of packets queuing at and the total bid prices from the wireless nodes. A decentralized learning algorithm incorporating a bidding mechanism is proposed, for scheduling separated information transmission and energy transfer, to obtain an optimal policy. The authors in \cite{D2014T} present a similar algorithm, based on decentralized online learning, to minimize the total number of packets queuing in the whole system. The authors also derive the conditions that the proposed algorithm converges to a local optimal solution. However, the authors ignore the bidding mechanism and consider random and unpredictable use of radio resource for information transmission and energy transfer.

\subsection{Receiver Operation Policy}

A receiver operation policy is required for wireless devices sharing the same antenna or antenna array for information reception and RF energy harvesting. The policy can be designed to deal with various tradeoffs in the physical layer and MAC layer to meet certain performance goals. Most of the existing policies are either based on \emph{time switching} or \emph{power splitting} architecture. The focus of the \emph{time switching} architecture is to coordinate the time for information reception and RF energy harvesting. By contrast, for the \emph{power splitting} architecture, the operation policy is to find an optimal ratio to split the received RF signals.

The majority of research efforts in receiver operation designs focus on a point-to-point single-input single-output (SISO) channel. Operation policies are proposed in fading channels without \cite{ZhangRuiMIMO,XZhou2013,LiangLiu,X1405.4623Zhou} 
and with co-channel interference \cite{Liu2012Liang,W1303.0381K,D2013W}. In \cite{ZhangRuiMIMO}, the authors investigate time-switching based policies with fixed or flexible transmit power constraints. The policies allow the signals transmitted to an information receiver and energy harvester to have the same fixed or different maximum power limit, respectively. The authors also analyze two power-splitting based policies, i.e., \emph{uniform power splitting} and \emph{antenna switching}. The \emph{uniform power splitting} assigns all the receiving antennas with the same power-splitting ratio. By contrast, the \emph{antenna switching} divides the total number of receiving antennas into two groups, each of which is either dedicated for information decoding or energy harvesting. 
The authors derive the achievable rate-energy regions for all the studied operation policies. 
The authors demonstrate that, when the RF band noise at the receiver antenna dominates the baseband processing noise, the power-splitting based policy approaches the upper bound of rate-energy region asymptotically.
It is also proved that the uniform power splitting outperforms the time-switching based policy with a fixed power constraint in terms of achievable rate-energy region. Nevertheless, the uniform power splitting is generally inferior to the time-switching based policy under the flexible power constraint without any peak power limit.

In~\cite{XZhou2013}, the authors generalize the time-switching and power-splitting based policies proposed in \cite{ZhangRuiMIMO} to a general dynamic power-splitting based policy. This policy dynamically splits RF signals into two streams with arbitrary ratio over time. The authors also investigate a special case of dynamic power-splitting, namely, the {\em on-off} power-splitting policy which divides the receiver into two modes. In the {\em off} mode, only the RF energy harvester receives all the RF signals. By contrast, in the {\em on} mode, the receiver operates as power-splitting. The analytical results prove that, for both co-located and integrated receiver architectures, the {\em on-off} power-splitting policy is optimal if taking receiver circuit power consumption into account. However, static power-splitting is optimal for the ideal case when the circuit power consumption is negligible.

The authors in \cite{X1405.4623Zhou} aim to achieve the best ergodic capacity performance for the power-splitting receiver through training. The system employs a block-wise transmission scheme, which consists of a training phase and a data transmission phase. The power-splitting ratios are optimally designed for both training and data phases to achieve the best ergodic capacity performance while maintaining required energy harvesting rate. The authors devise a non-adaptive and an adaptive power-splitting scheme in which the power-splitting ratios are fixed for all blocks, and are adjustable during data phases, respectively. It is proved that both schemes can provide optimal solutions.  The simulation demonstrates that the adaptive power-splitting design achieve a considerably improved capacity gain over the non-adaptive one, especially when the required energy harvesting rate is high.

The authors in \cite{Liu2012Liang} investigate the time-switching based operation policy for 
a fading channel with time-varying co-channel interference. For the case without CSI at transmitters, an opportunistic optimal time-switching policy, based on the instantaneous channel gain and interference power, is proposed to leverage the information and energy tradeoff as well as outage probability and energy tradeoff for delay-tolerant and delay-limited information transmission, respectively. By contrast, for the case with CSI at the transmitter, joint optimization of transmit power control with the receiver operation policy is also investigated. It is shown that for time-switching, an optimal operation policy is threshold-based. The policy instructs the user to harvest energy when the fading channel gain is above a certain threshold, and decode information otherwise. To reduce the complexity at the receiver, the authors devise heuristic operation policies, one of which performs time-switching based on co-channel interference. This policy is shown to achieve optimal performance. Besides, an insightful finding is that for opportunistic energy harvesting, the best policy to achieve the optimal information and energy tradeoff as well as outage probability and energy tradeoff is to allocate the fading states with the best channel gains to power transfer rather than information transfer. 

In \cite{LiangLiu}, the authors explore a power-splitting based operation policy for both SISO and single-input multi-output (SIMO) fading channels. It is found that, in a SISO channel, to achieve an optimal information and energy tradeoff for both cases with and without CSI at transmitters, the best policy for power-splitting based operation is to divide all the received signals according to a fixed ratio when the fading channel gain is above a certain threshold, and allocate all the signals to information receiver otherwise. Additionally, shown by simulations, the proposed power-splitting based policy can achieve substantial information and energy performance gains over the time-switching based policy proposed in \cite{Liu2012Liang}. Furthermore, the authors extend the power-splitting based policy to a SIMO channel, and show that the uniform power splitting policy is optimal. An antenna switching policy with low complexity is proposed and shown to achieve the performance close to that of the optimal uniform power splitting when the number of antennas at the receiver increases.

The study in \cite{W1303.0381K} considers the spectral efficient optimization problem in an OFDM-based system with a slow-fading channel and co-channel interference. A joint design of transmit power allocation and receiver operation based on power-splitting to maximize spectral efficiency of information transmission (i.e., bit/s/Hz) is formulated as a non-convex optimization problem. The optimal solution is obtained by a full search for the power-splitting ratio and convex optimal techniques. Then, two suboptimal iterative algorithms with low complexity are proposed to compromise between complexity and performance and shown to reach near optimal performance. Considering the same system model as in \cite{W1303.0381K}, the same authors also investigate the energy efficiency optimization problem in \cite{D2013W}. A similar joint design to maximize energy efficiency of information transmission (i.e., bit/Joule) is formulated as a multi-dimensional non-convex optimization problem. To solve the problem, the authors propose an iterative algorithm based on dual decomposition and a one-dimensional search. The simulation shows that the iterative algorithm can converge to an optimal solution. It is also revealed that system throughput maximization and energy efficiency maximization can be achieved simultaneously in the low transmit power regime. Besides, RF power transfer enhances the energy efficiency, especially in the interference limited regime.

The authors in \cite{H1301.4798Ju} introduce the time-switching based receiver operation policy for a point-to-point system with random beamforming applied at a multi-antenna transmitter. The theoretical analysis proves that, when transmit power approaches infinity, employing one single random beam enables the proposed policy to achieve an asymptotically optimal tradeoff between the average information rate and average harvested energy. The authors demonstrate by simulations that, even with finite transmit power, the proposed policy achieves the best information and energy/outage probability tradeoff with a single random beam employed for large power harvesting targets. 

Various operation policies have also been proposed for multi-user downlink systems with SISO channels \cite{Ng2013TWC,X1308.2462Zhou}, MISO channels \cite{Q1304.0062Shi}.
The investigation in \cite{Ng2013TWC} considers an OFDMA system with SWIPT from an energy efficiency perspective. Specifically, the joint design of power allocation, sub-carrier allocation and receiver operation policy is proposed to maximize system energy efficiency. In particular, the authors examine the case that the receivers can perform power splitting with arbitrary continuous splitting ratios, and the case that the receivers can only split the received signals into a discrete set of power streams with fixed splitting ratios. For each case, the authors formulate the joint design policy as a non-convex optimization problem and solve the problem by applying fractional programming and dual decomposition techniques. The simulation reveals that RF energy harvesting capability improves network performance in terms of energy efficiency, especially in the interference limited regime. 

The investigation in \cite{X1308.2462Zhou} considers receiver operation problems in multiple access channels with TDMA and OFDMA. The aim is to maximize the weighted sum-rate over all receivers under the constraints of minimum amount of harvested energy as well as peak and/or total transmit power. In particular, the authors propose a joint TDMA-based transmission with time-switching based policy as well as a joint OFDMA-based transmission with power-splitting based policy. Evaluated in a downlink OFDM system, it is proved that for a single-receiver case, the time-switching based policy outperforms the power-splitting based policy if there is no peak power constraint on each subcarrier. By contrast, the power-splitting based policy is superior when the peak power is sufficiently small. For a general multi-user case without the peak power constraint, it is shown numerically that the power-splitting based policy outperforms the time-switching based policy when the minimum required harvested energy is sufficiently small. However, with the finite peak power constraint, the time-switching based policy outperforms the power-splitting based policy when both the minimum harvested energy and the achievable rate are sufficiently large.

The studies in \cite{Q1304.0062Shi} and \cite{S1402.5730Leng} both investigate joint beamforming vector and receiver operation designs. 
In \cite{Q1304.0062Shi}, the authors present the joint beamforming vector and power-splitting ratio design for a multi-antenna base station. The objective is to minimize the total transmit power under the signal-to-interference-plus-noise ratio (SINR) and energy harvesting constraints at the receivers. The authors formulate this joint design as a non-convex problem and derive the sufficient and necessary condition for the feasibility of the problem. The semidefinite relaxation technique is applied and proved to achieve the globally optimal solution. Moreover, the authors propose two suboptimal solutions with low complexity, based on the zero-forcing and SINR-optimal criteria, respectively, for the formulated problem by designing the beamforming vectors and power-splitting ratio separately. The simulation results show that the two suboptimal solutions achieve comparable performance when SINR is greater than 5 dB, and perform very close to the optimal solution when SINR is larger than 20 dB.
The study in \cite{S1402.5730Leng} extends \cite{Q1304.0062Shi} by considering secure communication in presence of potential eavesdroppers. A non-convex problem is formulated to jointly optimize beamforming vectors, power splitting ratios and the covariance of the artificial noise, with an additional constraint on the maximum tolerable data rate at potential eavesdroppers. The problem is transformed into semidefinite programming and solved by semidefinite relaxation. The authors prove that the relaxation is tight and achieves the global optimum of the original problem.

The study in \cite{C1308.2838Shen} deals with the receiver operation problem in the system consisting of multiple transmitter-receiver pairs with MISO interference channels.  
The objective of \cite{C1308.2838Shen} is to maximize the system throughput subject to individual energy harvesting constraints and transmit power constraints. The authors first propose two time-switching based policy, namely, time division mode switching and the time division multiple access. The former divides each time slot into two sub-slots. All receivers harvest energy in the first sub-slot and subsequently decode information in the second sub-slot. The latter divides each time slot into $K$ sub-slots, and allows each receiver to take turn to decode information while the others harvest energy in each sub-slot. The authors further study an ideal-receiver based policy and a power-splitting based policy. The optimization problems associated with the two time-switching based policies are formulated as convex problems, while those with the ideal policy and the power-splitting based policy are solved by an approximation method based on log-exponential reformulation and successive convex approximation. An interesting observation is that the ideal policy may not always yield the best information and energy tradeoff compared with the proposed simple time-switching based policies. This is because the interference, which is avoided in time-switching based policies, degrades the achievable information rate for the ideal policy and power-splitting based policy. Another finding is that higher cross-link channel powers can increase the system throughput under energy harvesting constraints, which is different from the case in conventional interference channel without RF energy harvesting. This is due to the fact that interference helps to improve the amount of harvested energy, and thus indirectly benefits information transmission.

Both references \cite{Sai2014Zhao} and \cite{B2014Koo} deal with antenna-switching policy in MIMO channels. In \cite{Sai2014Zhao}, a MIMO downlink system with an information receiver and an energy harvester is studied. To maximize the achievable rate at the information receiver subject to the energy-harvesting constraint at the energy harvester and the transmit power constraint, a joint antenna-switching and transmit covariance matrix optimization is formulated as a non-convex mixed integer programming. An iterative antenna-switching algorithm is proposed to optimize the antenna switching at both the transmitter and information receiver as well as the transmit power over the selected antennas. Moreover, the authors propose a low-complexity
non-iterative norm-based algorithm which optimizes the antenna switching and transmit power sequentially. The simulation results show that the achievable rates of the proposed iterative algorithms approach that of the antenna-switching scheme optimized by an exhaustive search.
The study in \cite{B2014Koo} considers multiple transmitter-receiver pairs with MIMO interference channels.
The authors investigate the performance of the random antenna switching policy selecting antennas independently of channel realizations, and derive ergodic rate and expected harvested energy in closed-form. Moreover, an exhaustive search and an iterative receive antenna switching policy are proposed to explore the information and energy tradeoff. The evaluation shows that the antenna switching policy outperforms the uniform power splitting policy in terms of rate-energy region.    

Table~\ref{tab:operation} presents the summary of the existing receiver operation policies for RFEHNs. Time-switching, power-splitting and separated receiver architecture have been the mostly studied that existing policies are based on. We find that the point-to-point MIMO/SIMO channel and multi-user SIMO downlink system have not been investigated. Moreover, few existing receiver operation policies (e.g., \cite{Liu2012Liang}, \cite{C1308.2838Shen} and \cite{B2014Koo}) take co-channel interference into account. Dealing with interference would be a crucial concern for the practical designs of the future receiver operation policies.

\begin{table*}\footnotesize
\centering
\caption{\footnotesize Comparison of Receiver Operation Policies for RFEHNs.} \label{tab:operation}
\begin{tabular}{|p{1.2cm}|p{2.5cm}|p{3cm}|p{2.2cm}|p{3cm}|p{3cm}|} 
\hline
\footnotesize {\bf Literature} & {\bf Receiver architecture} & {\bf System model} & {\bf Channel model} & {\bf Design objective} & {\bf CSI requirement}    \\ \hline 
\hline
C. Shen \emph{et al} \cite{C1308.2838Shen}   & Time-switching, power-splitting & Multiple transmitter-receiver pairs with MISO channels & Cross-link interference channel  & Maximizing system throughput & N.A. \\
\hline
X. Zhou \emph{et al} \cite{X1308.2462Zhou}   & Time-switching, power-splitting  & Multi-user OFDM-based SISO downlink system & Slow-fading channels  & Maximizing weighted sum-rate over all receivers & N.A. \\
\hline
B. Koo \emph{et al} \cite{B2014Koo}   & Separated receiver architecture (Antenna-switching) & Multiple transmitter-receiver pairs with MIMO channels  &  Cross-link interference channel   &  Information and energy tradeoff & N.A. \\
\hline
S. Zhao \emph{et al} \cite{Sai2014Zhao}   & Separated receiver architecture (Antenna-switching) & MIMO downlink system with an information receiver and an energy harvester & AWGN channel & Maximizing throughput of information receiver & Perfect CSI at transmitter and receiver\\
\hline 
D. W. K. Ng \emph{et al} \cite{Ng2013TWC}   & Power-splitting & Multi-user OFDMA SISO downlink system &   Quasi-static block fading channels & Maximizing energy efficiency of information transmission  & Perfect CSI at receivers  \\
\hline
Q. Shi \emph{et al} \cite{Q1304.0062Shi}  & Power-splitting & Multi-user MISO downlink system & Quasi-static flat-fading channel &  Minimizing total transmit power  & N.A. \\
\hline
S. Leng \emph{et al} \cite{S1402.5730Leng}   & Power-splitting &  Multi-user MISO/MIMO downlink system  & Flat fading channel & Minimizing total transmit power  & Perfect CSI at transmitter \\
\hline
H. Ju \emph{et al} \cite{H1301.4798Ju}   & Time-switching & Point-to-point MISO system  & Quasi-static flat fading  & Information and energy tradeoff & None \\
\hline
X. Zhou \emph{et al} \cite{XZhou2013}   &  Time-switching, power-splitting, integrated & Point-to-point SISO system  & AWGN channel & Information and energy tradeoff & Perfect CSI at receiver \\
\hline
X. Zhou \emph{et al} \cite{X1405.4623Zhou} & Power-splitting  & Point-to-point SISO system & Block-wise Rayleigh fading & Maximizing ergodic capacity  & None \\
\hline
L. Liu \emph{et al} \cite{LiangLiu}   & Power-splitting & Point-to-point SISO/MISO channel & Flat fading channel & Information and energy tradeoff  & With and without CSI at transmitter \\
\hline
L. Liu \emph{et al} \cite{Liu2012Liang}   & Time-switching & Point-to-point SISO interference channel & Flat fading channel &  Information and energy tradeoff, outage probability and energy tradeoff  & With and without CSI at transmitter  \\

\hline
\end{tabular}
\end{table*}

\section{Multi-antenna RF-EHNs}
 
\begin{table*}\footnotesize
\centering
\caption{\footnotesize Summary of SWIPT Beamforming Designs in Multi-antenna RF-EHNs.} \label{SWIPT_Beamforming}
\begin{tabular}{|p{1.2cm}|p{2cm}|p{2.5cm}|p{2.8cm}|p{1.7cm}|p{3.2cm}|p{1.8cm}|} 
\hline
\footnotesize {\bf Literature} & {\bf Network model} & {\bf Design goal} & {\bf Constraints} & {\bf Problem formulation} & {\bf Solutions} & {\bf CSI requirement} \\ \hline 
\hline
J. Park \emph{et al} \cite{Park2013Jaehyun}  & Two transmitter-receiver pairs with MIMO interference channels  & Optimal transmission strategy for different cases & Transmit power & \textbf{P1} and \textbf{P2}: Linear program; \textbf{P3}: Convex optimization programming & \textbf{P1}: iterative water-filling;  \textbf{P2}: singular value decomposition; \textbf{P3}: Iterative algorithm based on singular value decomposition and subgradient-based method & Perfect CSI at transmitter \\
\hline
J. Park \emph{et al}  \cite{J1303.1693Park}   & Multiple transmitter-receiver pairs with MIMO interference channels  & Optimal transmission strategy for different cases & Transmit power & Non-convex programming& An iterative algorithm based on singular value decomposition and subgradient-based method & Perfect CSI at transmitter  \\
\hline
R. Zhang \emph{et al}  \cite{ZhangRuiMIMO}  & MIMO downlink system with a single information receiver and energy harvester  & {\bf P1}: Maximization of harvested energy; {\bf P2} and {\bf P3}: Maximization of information rate & {\bf P1}: Average transmit power; {\bf P2}: Average transmit power; {\bf P3}: Average transmit power and harvested energy requirement &  {\bf P1}: Linear programming; {\bf P2}: Convex programming; {\bf P3}: Convex programming & {\bf P1}: Singular value
decomposition; {\bf P2}: Singular value
decomposition and water filling algorithm; {\bf P3}: Lagrange duality method and singular value
decomposition & Perfect CSI at transmitter  \\
\hline
Z.~Xiang \emph{et al} \cite{Zhengzheng2012} & MISO downlink system with a single information receiver and energy harvester & Maximization of the worst-case harvested energy at energy harvester & Information rate target at information receiver & Semi-infinite non-convex quadratically constrained quadratic program & Semidefinite relaxation & Imperfect CSI at transmitter\\
\hline
S. Timotheou \emph{et al} \cite{S.Timotheou} & Multiple transmitter-receiver pairs with MISO interference channels & Minimization of total transmit power & Individual SINR and energy harvesting constraints at receivers & Non-convex quadratically constrained quadratic program & Semidefinite programming with rank relaxation & Perfect CSI at transmitters \\
\hline
H. Zhang \emph{et al} \cite{H.2013Zhang} & MISO downlink  system with multiple information receivers and energy harvesters & Maximization of the amount of energy harvested at the worst energy harvester & Total transmit power limit and SINR requirements at information receivers & Non-convex programming & Semidefinite relaxation  & Imperfect CSI at transmitter\\
\hline
J. Xu \emph{et al} \cite{J1303.1911Xu} & MISO downlink system with multiple information receivers and energy harvesters & Maximization of weighted sum energy transferred & Individual SINR constraints at information receivers  & Non-convex quadratically constrained quadratic program & Semidefinite relaxation, uplink-downlink duality & N.A. \\
\hline
M. R. A. Khandaker \emph{et al} \cite{Muhammad2014R} & Downlink MISO multicasting system with multiple power-splitting receivers & {\bf P1} and {\bf P2:} Minimization of total transmit power of base station  & {\bf P1} and {\bf P2:} SNR and energy
harvesting constraints at each receiver & {\bf P1} and {\bf P2:} Non-convex problem & {\bf P1:} Semidefinite relaxation, Hermitian matrix rank-one decomposition
techniques; {\bf P2:} Semidefinite relaxation, interior point methods & {\bf P1:}  Perfect CSI at transmitter; {\bf P2:}  Imperfect CSI at transmitter\\
\hline
D. Li \emph{et al} \cite{DLi2013} & Analog network coding based two-way multiple-relay system & Maximization of weighted sum rate & Transmit power limit at relays and energy harvesting requirement at sources & \textbf{P1} and \textbf{P2}: Non-convex Programming & \textbf{P1}: Semidefinite relaxation and successive convex approximation; \textbf{P2}: Semidefinite relaxation, Charnes-copper transformation and successive convex approximation  & N.A. \\
\hline
D. Li \emph{et al} \cite{D2013Li} &  AF based two-way multiple relay system & Maximization of weighted sum rate &  Transmit power limit at relays and energy harvesting requirement at sources &  Non-convex programming  & Semidefinite relaxation, S-procedure and successive convex approximation  & Imperfect CSI at relays  \\
\hline 
D. W. K. Ng \emph{et al} \cite{D1403.5730W} & Multi-user coordinated multipoint network with SWIPT  & Jointly minimization of total transmit power and maximum capacity consumption per backhaul link  &  Minimum required SINR constraint at information receivers and minimum harvested energy constraint at energy harvesters & Non-convex programming & Semidefinite relaxation, a local-optimal iterative algorithm  & Perfect CSI at central processor \\ 
\hline 
J. Park  \emph{et al} \cite{J1403.2189Park} & Information transmitter-receiver pair and an energy  transmitter-receiver pair & Joint maximization of amount of harvested energy at energy harvester and minimization of interference to information receiver & Energy harvesting constraint at energy harvester and rank-one constraint on transmit signal covariance of information transmitter & Convex problem & Optimal Geodesic information/ energy beamforming schemes & Partial CSI at energy/information transmitters \\
\hline

\end{tabular}
\end{table*}
 
\begin{figure}
\centering
\includegraphics[width=0.4\textwidth]{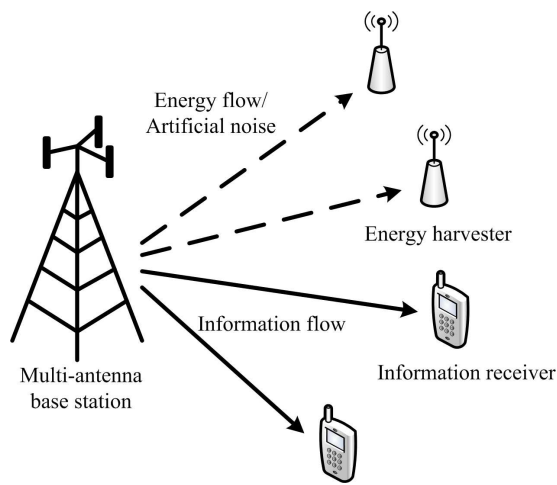}
\caption{A general model for SWIPT beamforming system.} \label{beamforming}
\end{figure} 
 
A key concern for RF information and energy transfer is the decay in energy transfer efficiency with the increase of transmission distance due to propagation path loss. This problem is especially severe in a single-antenna transmitter which generates omni-directional radiation of emitted RF signals. 
The low energy transfer efficiency of RF energy calls for advanced multi-antenna  and signal processing techniques such as beamforming. Multi-antenna techniques can  achieve spatial multiplexing. Furthermore, beamforming techniques employing multiple antennas can be applied to improve efficiency of RF energy transfer \cite{X2013Chen} as well as SWIPT \cite{ZhangRuiMIMO}, without additional bandwidth or increased transmit power. Indeed, beamforming has been deemed as a primary technique for feasible implementation of SWIPT \cite{O2002Mcspadden,K2013Huang,C1984Brown}.

\begin{figure}
\centering
\includegraphics[width=0.4\textwidth]{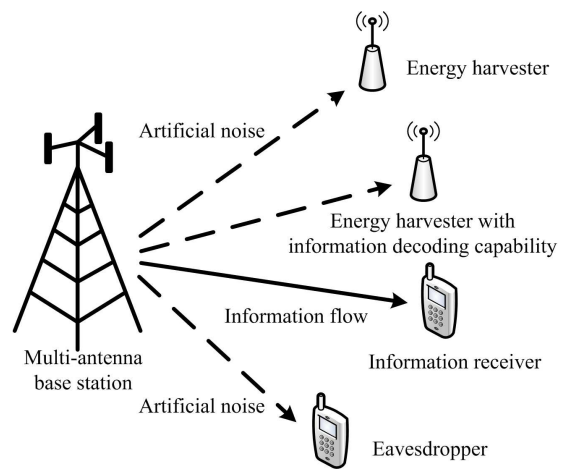}
\caption{A model for SWIPT beamforming with secure communication.} \label{beamforming_secure}
\end{figure} 

In RF-EHNs, beamforming designs have been explored to steer the RF signals toward the target receivers with different information and/or energy harvesting requirements. Figure~\ref{beamforming} shows a general model for the SWIPT beamforming system. In addition to data transmission and energy harvesting optimization for beamforming, existing literature has also exploited beamforming to ensure secure communication. The security issue in RF-EHNs with SWIPT is that, a transmitter may amplify the power of information transmission to facilitate energy harvesting at receivers. Consequently, it may result in more vulnerability to information leakage. Therefore, beamforming with secrecy requirement has to be  designed taking into account potential eavesdropping.
Recent work has advocated the dual use of  artificial noise/interference signals in facilitating RF energy transfer and providing security for information transmission.
The idea is to impair the received signals at potential eavesdroppers (e.g., unauthorized information receivers and energy harvesters with information decoding capability) by injecting artificial noise into their communication channels. Figure~\ref{beamforming_secure} shows a network model for SWIPT beamforming with secure communication.

The knowledge of CSI plays an important role in beamforming performance optimization. To accurately estimate a channel state, a significant overhead (e.g., time) can be incurred at a receiver. Normally, the longer time for channel state estimation contributes to more accurate CSI, which, however, results in reduced time for transmission, and also less amount of harvested energy. As a result, an optimization of RF energy transfer or SWIPT entails a tradeoff between data transmission and channel state estimation duration. Another problem arising in beamforming is channel state estimation feedback. Designing a feedback mechanism is challenging in RF-EHNs because existing channel training and feedback mechanisms used for an information receiver are not applicable for an energy harvester due to the hardware limitation, i.e., without baseband signal processing.

This section reviews the beamforming designs for RF-EHNs. We classify the related work into three categories, i.e., SWIPT beamforming without and with secure communication requirement as well as energy beamforming designs. The feedback mechanisms for beamforming  are also included. 

\subsection{SWIPT Beamforming Optimization without Secure Communication Requirement}

Beamforming is first explored in a three-node MIMO network~\cite{ZhangRuiMIMO} with one transmitter, one energy harvester and one information receiver. The authors study the optimal transmission strategies to achieve tradeoff between information rate and the amount of RF energy transferred under the assumption of perfect knowledge of CSI at the transmitter. The study in \cite{Zhengzheng2012} extends \cite{ZhangRuiMIMO} by considering imperfect CSI at the transmitter in a three-node MISO network. The objective is to maximize the worst-case amount of harvested energy at the energy harvester with the information rate target for all possible channel realizations. Additionally, the authors consider a robust beamforming design. As both the expressions of harvested energy and the information rate constraint are quadratic, the design  is modeled as a semi-infinite non-convex quadratically constrained quadratic programming problem with infinite constraints due to the channel uncertainties. By applying semidefinite relaxation, the original problem is then transformed into a convex semidefinite program and it can be solved efficiently. The theoretical proof indicates that the solution of the relaxed semidefinite program is always rank-one.

Both references \cite{H.2013Zhang} and \cite{J1303.1911Xu} consider a MISO downlink broadcast system with multiple separated information receivers and energy harvesters. In \cite{H.2013Zhang}, the authors aim to maximize the amount of energy harvested at the energy harvester that collects the least amount of energy. To achieve the goal, the authors present a robust beamforming design considering imperfect CSI at the transmitter under the constraints of SINR at information receivers and total transmit power. As the considered design problem is non-convex in general, it is converted to the relaxed semidefinite problem and solved by standard interior-point methods. The authors further propose an iterative algorithm based on the bisection method to obtain the robust beamforming solution, the performance of which is demonstrated to be very close to the upper bound. The objective of \cite{J1303.1911Xu} is to maximize the weighted sum energy transferred to energy harvesters while guaranteeing individual SINR requirements at information receivers. In particular, two types of information receivers without and with the capability of canceling the interference are studied. For each type of the information receiver, the authors formulate the joint transmit beamforming weight and power allocation design as a non-convex quadratically constrained quadratic programming problem. The authors obtain the globally optimal solutions for the formulated problem by means of semidefinite relaxation. It is shown that the solutions of the relaxed semidefinite program for both types of information receivers are rank-one. Moreover, it requires no dedicated energy beam to achieve the optimal solution for information receivers without interference cancellation. The authors also design an alternative approach based on uplink-downlink duality to obtain the same optimal solutions by semidefinite relaxation. 

Different from \cite{H.2013Zhang} and \cite{J1303.1911Xu}, in \cite{Muhammad2014R}, multiple power-splitting receivers are considered in a MISO downlink broadcast system. The authors investigate the joint design of multicast transmit beamforming and receiver power-splitting ratio for minimizing the transmit power of a base station with SNR and energy harvesting constraints at each receiver. For the cases with perfect or imperfect CSI, a non-convex programming problem is formulated and solved applying semidefinite relaxation techniques. The conditions when the relaxation is tight are also derived.

The study in \cite{S.Timotheou} deals with a total transmit power minimization problem for a MISO network consisting of multiple transmitter-receiver pairs with co-channel interference. For each of the considered fixed beamforming schemes, i.e., zero-forcing (ZF), regularized ZF, maximum ratio transmission (MRT), and a linear combination of ZF and MRT, called MRT-ZF, a joint design of transmit power allocation and receiver power-splitting ratio is formulated as an optimization problem assuming power-splitting architecture adopted at the receivers.
It is found that MRT significantly outperforms ZF in terms of transmit power, because the co-channel interference, which is canceled in ZF beamforming, is beneficial for energy harvesting in MRT beamforming. However, the side-effect for MRT beamforming is that it may result in infeasible solutions while ZF beamforming ensures existence of feasible solutions. The reason is because, without co-channel interference, ZF beamforming requires more energy to achieve desired level of energy harvesting which also improves SINR. Regularized ZF is not suitable for the considered problem as it exhibits the most infeasibility issue. By combining the best of MRT and ZF, MRT-ZF always results in feasible solutions with considerable better performance than those of the other schemes. The authors further investigate adaptive beamforming designs in the transmit power minimization problem. A joint design of beamforming weights, transmit power and power-splitting ratio is formulated and solved by semidefinite programming with rank relaxation. Moreover, a heuristic algorithm is proposed to obtain the beamforming solution when the rank relaxation is not tight (i.e., an optimal solution is not available). It is theoretically proved that the proposed approach always gives rank-one solutions when there are two or three transmitter-receiver pairs. 

Beamforming design problems are studied in \cite{DLi2013} and \cite{D2013Li} for a two-way relay system. Two single-antenna source nodes exchange information through multiple relay nodes, and harvest RF energy from the transmission of these relays. The objective of \cite{DLi2013} is to maximize weighted sum-rate with the transmit power limit and energy harvesting constraints. Under the assumption of an ideal receiver architecture at the source nodes, an iterative algorithm based on semidefinite relaxation and successive convex approximation is devised to obtain a local optimal solution. Then, the iterative algorithm is extended to the case of the time-switching architecture at the source nodes. The authors in \cite{D2013Li} extend the case with the ideal receiver architecture in \cite{DLi2013} by exploiting imperfect CSI at relays. A robust beamforming design problem is formulated to optimize the same objective assuming amplify-and-forward (AF) based relays. To handle infinity of the non-convex constraints due to channel uncertainty, the authors first reformulate the optimization problem by approximating the SINR. Then, semidefinite relaxation, S-procedure and successive convex approximation techniques are applied to address the reformulated problem. 

The authors in \cite{Park2013Jaehyun} and \cite{J1303.1693Park} investigate SWIPT beamforming in a MIMO system consisting of multiple transmitter-receiver pairs with co-channel interference, where all the receivers adopt time-switching architecture. The study in \cite{Park2013Jaehyun} focuses on the case with two transmitter-receiver pairs. For the cases when both receivers work as information receivers or energy harvesters, the authors study the achievable rate for an iterative water-filling algorithm without CSI sharing between two transmitters. The authors also devise an optimal strategy based on singular value decomposition to maximize transferred energy. For the case when one adopts an information receiver and the other adopts an energy harvester, the authors develop two rank-one beamforming strategies with the objective to maximize transferred energy to the energy harvester and minimize interference to the information receiver. Furthermore, the performance metric, called signal-to-interference-and-harvested, is introduced to maximize transferred energy as well as minimize interference to the information receiver simultaneously. A rank-one beamforming strategy developed based on this metric is shown to achieve a wider rate-energy region than those of the other two proposed beamforming strategies. The authors in \cite{J1303.1693Park} generalize the problem investigated in \cite{Park2013Jaehyun} to the case of $k$ transmitter-receiver pairs. The different scenarios are considered, i.e., multiple energy harvesters/information receivers and a single information receiver/energy harvester as well as multiple information receivers and multiple energy harvesters. For all the scenarios, the authors define necessary conditions for optimal energy transmitters. Accordingly, the transmission strategies that exploit rank-one beamforming at energy transmitters are developed. The three beamforming strategies proposed in \cite{Park2013Jaehyun} are modified to be applicable for the considered case. Moreover, to derive achievable rate-energy region given that the energy transmitters adopt rank-one beamforming, the authors formulate the non-convex optimization problem and propose an iterative algorithm to solve it. An interesting finding is that when the number of information transmitters increases, the optimal beamforming strategy approximates the beamforming strategy that maximizes harvested energy. This is because the interference from information transmission is beneficial for energy harvesting. 

The focus of \cite{J1403.2189Park} is to reduce the feedback overhead in a two-user MIMO channel. An access point serves an information receiver and a power charger serves an energy harvester by sharing the same spectrum resource. The authors propose a Geodesic energy beamforming scheme and a Geodesic information beamforming scheme that require only partial CSI at both power charger and access point. The authors prove that the Geodesic information/energy beamforming approach is an optimal strategy for SWIPT in the two-user MIMO interference channel, under the rank-one constraint of transmit signal at the access point.

The beamforming design for coordinated multi-point networks with SWIPT is addressed in \cite{D1403.5730W}. The system under consideration contains a central processor, which is connected to the transmitters via capacity-limited backhaul links, to facilitate coordinated multi-point transmission. The objective is to jointly minimize the total network transmit power and maximum capacity consumption per backhaul link, with the minimum SINR and harvested energy constraints at the information receivers and  energy harvester, respectively. However, this results in a non-convex programming problem. The authors propose a suboptimal iterative algorithm and prove that it can obtain a locally optimal solution. The simulation results show that the proposed scheme performs close to the optimal scheme based on an exhaustive search. The study also shows the potential power savings enabled by coordinated multi-point networks compared to centralized multi-antenna systems.

Table~\ref{SWIPT_Beamforming} shows the summary of the above reviewed SWIPT beamforming designs.     
    
\subsection{SWIPT beamforming for Secure Communication}

\begin{table*}\footnotesize
\centering
\caption{\footnotesize Summary of SWIPT Beamforming Designs for Secure Communication in Multi-antenna RF-EHNs.} \label{SWIPT_Secure}
\begin{tabular}{|p{1.2cm}|p{2cm}|p{2.5cm}|p{2.8cm}|p{1.5cm}|p{2.5cm}|p{2cm}|} 
\hline
\footnotesize {\bf Literature} & {\bf Network model} & {\bf Design goal} & {\bf Constraints} & {\bf Problem formulation} & {\bf Solutions} & {\bf CSI requirement} \\ \hline 
\hline
D. W. K. Ng \emph{et al} \cite{Ng1311.2507} & A MISO downlink system with a targeted and multiple idle information receivers  & Minimization of transmit power & Different SINR constraints at desired receiver and potential eavesdroppers, outage probability constraint at the passive eavesdroppers, and energy harvesting constraint at idle legitimate receivers  & Non-convex programming problem & Semidefinite relaxation & Imperfect CSI of potential eavesdroppers and no CSI of passive eavesdroppers at receiver
 \\
\hline
L. Liu \emph{et al} \cite{L1307.6110Liu} & A MISO downlink system with a single information receiver and multiple energy harvesters & \textbf{P1}: Maximization of secrecy information rate; \textbf{P2}: Maximization of weighted sum harvested energy & \textbf{P1}: individual harvested energy constraints; \textbf{P2}: secrecy information rate constraint & \textbf{P1} and \textbf{P2}: non-convex programming problem & \textbf{P1} and \textbf{P2}: Semidefinite relaxation and one-dimension search & N. A. \\
\hline
B. Zhu \emph{et al} \cite{B2014Zhu} & A MISO downlink system with multiple information receivers and energy harvesters &  Minimization of total transmit power & SINR and energy harvesting constraints at information receivers and energy harvester, respectively & Quadratically constrained quadratic
program & Rank-two beamformed Alamouti coding and semidefinite relaxation, a rank-two Gaussian randomization procedure  & Perfect CSI at transmitter \\
\hline
D. W. K. Ng  \emph{et al} \cite{Ng1309.2143}  & A MISO downlink system with multiple targeted and idle information receivers as well as passive eavesdroppers & Minimization of the total transmit power & Heterogeneous QoS requirements for multicast video receivers and energy harvesting requirements at idle receivers & Non-convex programming problem & Semidefinite relaxation, two sub-optimal algorithms  & Perfect CSI at receivers \\
\hline
Q. Shi \emph{et al} \cite{Q1403.31960Shi}  &  A MIMO downlink system with an information receiver and an  eavesdropping energy harvester & Maximization of secrecy information rate  & Harvested power constraint and the total transmit power constraint at the RF-powered information receiver and transmitter, respectively & Non-convex programming problem & Semidefinite relaxation, eigen-decomposition, rank-one reduction technique, an inexact block coordinate descent algorithm & N. A. \\
\hline
D. W. K. Ng \emph{et al} \cite{D1403.0054W} & A downlink MISO secondary communication system & Joint maximization of energy harvesting efficiency, and minimization of total transmit power and interference power leakage to transmit power ratio & SINR constraints at secondary information receiver, eavesdropper and primary network, transmit power constraint at secondary transmitter  & Non-convex programming & Semidefinite relaxation & Imperfect CSI at idle secondary receivers and primary receivers  \\
\hline
Q. Li \emph{et al} \cite{Q2014Li} & A two-hop single relay system with an information receiver, an energy harvester and an eavesdropper  &  Maximization of secrecy information rate & Relay transmit power constraint and energy harvesting requirement at the energy harvester & Non-convex programming problem &  An iterative algorithm based on constrained concave convex procedure & Perfect CSI at the relay \\
\hline
\end{tabular}
\end{table*}

The authors in \cite{L1307.6110Liu} 
and  \cite{Ng1311.2507}  investigate the beamforming schemes for secure communication in MISO downlink systems with a single targeted information receiver and other idle information receiver(s) or energy harvester(s) which can be eavesdropper(s). 
In \cite{L1307.6110Liu}, the authors investigate a joint design of transmit beamforming vectors and power allocation with different objectives.
Specifically, the former aims to maximize the secrecy rate for the information receiver under individual harvested energy constraints of energy harvesters. The latter maximizes the weighted sum harvested energy at the energy harvesters with the secrecy rate constraint for the information receiver. Both are formulated as non-convex problems and solved by a two-stage optimization approach based on the semidefinite relaxation and one-dimension search. 
Furthermore, two suboptimal solutions with low complexity, which design the beamforming vectors separately with power allocation, are proposed for each of the studied problems. The first suboptimal solution attempts to eliminate the information leakage by aligning the information beam to the null space of the energy harvesters. The second suboptimal solution aligns the information beams to the same direction to maximize SINR at the information receiver. The simulation results show that the second suboptimal solution achieves better information and energy tradeoff at the cost of higher complexity. 
The authors in \cite{Ng1311.2507} consider a secure communication guarantee via artificial noise injection, with imperfect CSI of potential eavesdroppers and no CSI of passive eavesdroppers. The CSI uncertainty introduces a non-convex probabilistic constraint in the formulated transmit power minimization problem.  
To tackle this issue, the authors replace the
non-convex probabilistic constraint with a convex deterministic constraint, and adopt semidefinite relaxation to obtain the optimal solution. It is found that the energy harvesting efficiency improves with the number of receivers, however, at the cost of higher transmit power.

Although some of the above works such as \cite{L1307.6110Liu,Ng1311.2507} design secure transmit beamforming based on rank-one semidefinite relaxation, they are only applicable for a single information receiver scenario. In \cite{B2014Zhu} and \cite{Ng1309.2143}, the authors study a more general system model with secure information multi-casting (i.e., with multiple information receivers). In \cite{B2014Zhu}, instead of utilizing artificial noise, the authors propose to use rank-two beamformed Alamouti space-time coding \cite{X2012Wu}
to develop secure multicast design for SWIPT. 
Specifically, a secure multicast design employing rank-two beamformed Alamouti coding, and semidefinite relaxation is proposed to address the total transmit power minimization problem under SINR and energy harvesting constraints at the information receivers and energy harvester, respectively.   
The authors derive sufficient conditions under which the rank-two semidefinite relaxation design is tight, and propose a rank-two Gaussian randomization procedure to obtain a suboptimal solution when the semidefinite relaxation design is not tight. In \cite{Ng1309.2143}, the authors study information multicasting in a TDMA-based secure layered transmission system consisting of multiple information receivers in presence of passive eavesdroppers. The aim is to design a power allocation method that minimizes the total transmit power accounting for the energy harvesting requirement at idle receivers and heterogeneous QoS requirements for multicast video receivers. As this design is shown to be an intractable non-convex optimization problem, it is reformulated by introducing a convex deterministic constraint. The authors develop semidefinite relaxation based power allocation to obtain the upper bound solution for the reformulated problem. Furthermore, two subptimal power allocation schemes are devised and demonstrated to give near optimal performances.

The study in \cite{Q1403.31960Shi} deals with the beamforming design in MIMO broadcast systems.
The authors consider a simple three-node network consisting of a transmitter, an intended information receiver and an energy harvester that can eavesdrop. A beamforming design is formulated as a non-convex problem to maximize the secrecy information rate subject to the transmit power constraint and energy harvesting constraint. The authors derive an equivalent problem of the beamforming design and propose an inexact block coordinate descent algorithm to obtain the solution. It is proved that the proposed algorithm can monotonically converge to the Karush-Kuhn-Tucker solution of the formulated problem. The authors also show that the proposed algorithm can be extended to a joint beamforming design and artificial noise generation problem.


The work in \cite{D1403.0054W} explores the beamforming design in a cognitive radio network. With the objective to jointly maximize energy harvesting efficiency and minimization of both total transmit power and interference power leakage-to-transmit power ratio, a multi-objective non-convex programming problem is formulated and recast as a convex optimization problem via semidefinite relaxation. By exploiting the primal and dual optimal solutions of the relaxed problem, the globally optimal solution of the original problem can be obtained. The authors further devise two suboptimal schemes for the case when the solution of the dual problem is not available. It can achieve the near-optimality of the suboptimal schemes.

Different from the above literature, the investigation in \cite{Q2014Li} copes with the beamforming design in a two-hop relay network. Specifically, the objective is to maximize the secrecy rate of a non-generative multi-antenna relay forwarded to an information receiver, subject to the transmit power constraint and  energy harvesting requirement of an energy harvester in presence of an eavesdropper. Under the assumption that CSI of the whole system is available at the relay, the authors introduce an iterative algorithm based on the constrained concave convex procedure, which is proved to achieve a local optimum. To ease computation complexity, the authors also propose a semidefinite relaxation based non-iterative suboptimal algorithm and a closed-form suboptimal algorithm. The simulation results illustrate that when SNR is high, the semidefinite relaxation based non-iterative suboptimal algorithm performs close to the proposed iterative algorithm.

Table~\ref{SWIPT_Secure} shows the summary of SWIPT beamforming designs with secure communication.     

\subsection{Energy Beamforming}

\begin{table*}\footnotesize
\centering
\caption{\footnotesize Summary of Energy Beamforming Designs for Multi-antenna RF-EHNs.} \label{Energy_beamforming}
\begin{tabular}{|p{1.2cm}|p{3.5cm}|p{2.1cm}|p{3.2cm}|p{1.5cm}|p{2.4cm}|p{1.2cm}|} 
\hline
\footnotesize {\bf Literature} & {\bf Network model} & {\bf Design goal} & {\bf Constraints} & {\bf Problem formulation} & {\bf Solutions} & {\bf CSI requirement} \\ \hline 
\hline
G. Yang \emph{et al} \cite{G1311.4111Yang}   & A point-to-point MISO system & Maximization of total harvested energy & Transmit power limit & Dynamic programming & Threshold-type policy & Imperfect CSI at energy transmitter \\
\hline
Q. Sun \emph{et al} \cite{Q1403.4492Sun} & A multi-user MISO system with a dedicated power station &  Maximization of the system sum-throughput & Transmission time constraint and transmit power constraint at the each user and the power station, respectively & Non-convex programming & Semidefinite relaxation & Perfect CSI at energy transmitter \\
\hline
X. Chen \emph{et al} \cite{X2013Chen} & A time division duplex large-scale MIMO system with a co-located energy transmitter and information receiver as well as an information transmitter & Maximization of system energy efficiency & Constraints of transmit power, sub-slot duration for RF energy transfer, and information rate & Fractional programming problem & Lagrange multiplier method, an iterative algorithm based on Dinkelbach method \cite{W1967Dinkelbach} & Perfect CSI at energy transmitter \\
\hline
S. Lee  \emph{et al} \cite{S1402.6441Lee} & Multiple transmitter-receiver pairs with SISO interference channels and a network coordinator & Optimization of rate-energy tradeoff & Energy harvesting constraints at receivers  & Non-convex programming  & Lagrange duality method & Perfect CSI at receivers  \\
\hline

\end{tabular}
\end{table*}

In \cite{G1311.4111Yang}, the authors design an adaptive energy beamforming scheme based on imperfect CSI feedback in a point-to-point MISO system. The considered system adopts a frame-based protocol, in which the receiver first performs channel estimation through the preambles sent by the transmitter and feeds the estimated CSI back to the transmitter. Then, the transmitter transmits via beamforming. The focus is to maximize the harvested energy by exploiting the tradeoff between channel estimation duration and power transfer duration as well as allocating transmit power. The authors first derive the optimal energy beamformers. Then, they obtain an optimal online preamble length and an offline preamble length, for the scenarios with variable and fixed length preambles, respectively. The transmit power is allocated according to both the optimal preamble length and the channel estimation power.

The studies in \cite{Q1403.4492Sun} and \cite{X2013Chen} investigate energy beamforming in multi-user systems. The authors in \cite{Q1403.4492Sun} consider a TDMA-based MISO system powered by a power station. A joint time allocation and energy beamforming design is formulated as a non-convex programming problem to maximize the system sum-throughput. The authors apply the semidefinite relaxation technique to reformulate the problem as a convex problem and prove the tightness as well as the global optimality of the semidefinite relaxation approximation. Furthermore, a fast semi-closed form solution is devised and shown by simulations to substantially reduce implementation complexity. 

In \cite{X2013Chen}, the authors deal with the resource allocation problem to improve energy efficiency of information rate (i.e., bit per Joule) in a large-scale MIMO system. The system consists of two components, i.e., a co-located energy transmitter and information receiver as well as an information transmitter. The system operates on a simple time-slot based scheme which divides each time-slot into two sub-slots. During the first sub-slot, the energy transmitter delivers RF energy to the information transmitter that  transmits information during the second sub-slot. A fractional programming problem, non-convex in general, is formulated to maximize energy efficiency taking into account the transmit power constraints, a time duration constraint for RF energy transfer, and an information rate requirement. The authors first use the Lagrange multiplier method to obtain the dual problem. Then, the resource allocation scheme based on a Dinkelbach method \cite{W1967Dinkelbach} is proposed to jointly optimize the transmit power and time duration of RF energy transfer. The proposed scheme is shown to have fast convergence speed and reach higher energy efficiency with the increased number of antennas. 

Different from the above work, the authors in \cite{S1402.6441Lee} exploit collaborative energy beamforming with distributed single-antenna transmitters. To facilitate the collaborative energy beamforming, a novel signal splitting scheme is introduced at the transmitters. For the case of two transmitter-receiver pairs, the authors propose a joint energy beamforming design with signal splitting to optimize the rate-energy tradeoff. For the case of more than two transmitter-receiver pairs, two suboptimal schemes of low complexity are devised. The first scheme divides all the pairs into different groups that contain two pairs and then applies the design for two-pair case directly. The other scheme is based on the \emph{ergodic interference alignment} technique \cite{B2012Nazer}, which requires synchronization for all receivers. The simulation shows that the latter scheme outperforms the former scheme due to interference-free degree of freedom.   

Table~\ref{Energy_beamforming} shows the summary of the existing energy beamforming designs. Compared to SWIPT beamforming, beamforming designs for dedicated energy transfer has been less investigated. Energy beamforming needs to be exploited in more diverse systems, such as heterogeneous networks.  

\subsection{Information Feedback Mechanism}

The study in \cite{J1312.1444Xu} aims to tackle the problem of information feedback for a practical energy harvester. Specifically, the authors devise a channel learning method for a transmitter to acquire CSI in the point-to-point MIMO network with RF energy beamforming. The method relies on one-bit information, which is a measurement of the increase or decrease of the amount of harvested energy at the energy harvester between the present and previous intervals, for feedback. Consequently, the energy transmitter can adjust transmit energy beamforming and obtain better estimation of the MIMO channel based on the feedback information. Compared with a cyclic Jacobi technique based method \cite{Y1301.2030Noam} and a stochastic gradient method \cite{C2003Banister}, the proposed learning method is shown to achieve lower normalized error and higher average harvested power.

The study in \cite{X2013Chenbeamforming} considers the information rate maximization problem for an information transmitter that harvests RF energy from a multi-antenna energy transmitter with beamforming. The authors propose to use quantization codebooks of limited size to provide feedback of CSI from the information transmitter to the energy transmitter for adaptive energy beamforming. For a given codebook size, the expressions of the upper bound and the approximate lower bound of the average information transmission rate are obtained. The expression reveals the relationship between the amount of CSI feedback with transmit power and transfer duration. Then, the authors introduce two schemes to optimize information and energy tradeoff for maximizing the upper bound and the approximate lower bound in a multi-antenna system with limited feedback. The performance impact of imperfect CSI is investigated, and the corresponding upper bound on the average information rate is derived.
 
In \cite{Y1403.7870Zeng}, the authors study the optimal design of an channel-acquisition scheme for a point-to-point MIMO energy beamforming system.  
Based on channel reciprocity, the energy transmitter estimates the CSI via dedicated reverse-link training from the energy harvester. A tradeoff in energy beamforming system is revealed between training time and energy transmission. In particular, too little training results in coarse CSI estimation and thus reduces energy beamforming gain. By contrast, too much training consumes excessive energy harvested by the energy harvester, and hence reduces the amount of time for energy transmission. To cope with the tradeoff, the optimal training design is proposed to maximize the net energy of the energy harvester, calculated by extracting energy used for channel training from the total harvested energy.

\section{Multi-hop RF-EHNs}   
   
In multi-hop relay networks, cooperative relaying techniques can help to overcome fading and attenuation by using intermediate relay nodes, resulting in improved network performance in terms of efficiency and reliability. Therefore, it is particularly suitable to be applied in energy constrained networks like RF-EHNs. For cooperative relaying in RF-EHNs, most research efforts attempt to improve performance gain on the physical layer and MAC layer (e.g., relay operation policy and power allocation) as well as network layer (i.e., relay selection). Other issues such as precoder maximization and cooperative scheme are also studied. These design issues in cooperative relaying become more complex for incomplete CSI. The information about energy status (e.g., energy reserve and potential available RF energy) must also be taken into account.  In the following, we review the related work of multi-hop RF-EHNs from different perspectives.

\subsection{Relay Operation Policy}

Section V has introduced the operation policies designed for receivers to achieve some information and energy tradeoff in different systems.
For relay nodes with RF energy harvesting capability in multi-hop networks, such operation is also required with additional consideration to the transmission requirement. 
The research works on the relay operation policy for relay nodes mainly consider a simple three-node cooperative relaying network composed of a source node, a relay node and a destination node. One or more of the nodes have the RF energy harvesting capability. All the communication traffic between source and destination node is assumed to be forwarded through the relay node. These schemes are studied based on common cooperative strategies, i.e., amplify-and-forward (AF) and decode-and-forward (DF). It is claimed in \cite{K2012IshibashiTWC} that AF may impose high peak power levels which makes DF scheme more practical, especially for energy constrained devices.

The authors in \cite{IKrikidis2012} examine a simple greedy switching policy based on the time-switching receiver architecture. The idea of the policy is to let the relay node transmit when its remaining energy can support information transmission. Based on the Markov chain model of the policy, the authors derive the closed-form expression of the outage probability for the relay node with a discrete-level battery. Compared with an optimal genie-aided policy that incorporates a priori knowledge of the channel coefficients and energy status of the relay node, the greedy switching policy is shown to reach the performance close to that of genie-aided policy in terms of outage probability over a wide range of SNR.

The study in \cite{I2014Krikidis} exploits the array configuration at a relay node in MIMO relay channels. The authors proposed two dynamic antenna switching policy which allocates a certain number of strongest channels for information decoding/energy harvesting and the remaining channels for energy harvesting/information decoding. The outage probability of the proposed policies is derived in closed-form.
The proposed policy is also analyzed in the scenarios with co-channel interference, where  the relay node adopts a zero-forcing receiver. The outage probability in closed-form expressions is derived.  
  
In \cite{SDurrani2013}, two relaying protocols for an RF energy harvesting relay node are proposed based on the \emph{time-switching} and \emph{power-splitting} receiver architectures, both of which require perfect CSI at the destination node. Specifically, the authors consider both the non-delay-limited and the delay-limited transmission, and derive the analytical expressions for the ergodic capacity and outage probability, respectively. The optimal RF energy harvesting time for the \emph{time-switching} based relaying protocol and the optimal value of power-splitting ratio for the \emph{power-splitting} based relaying protocol can be obtained. The evaluation results conclude that the \emph{time-switching} based relaying protocol is superior in terms of throughput at relatively low SNR and high transmission rates. However, as the transmit power is variable, it incurs significant hardware complexity. Consequently, the relay node may require a large dynamic range of the power amplifier \cite{A1310.7648Nasir}. 
  
Based on the time-switching receiver architecture, the authors in \cite{A1310.7648Nasir} propose adaptive time-switching protocols for RF energy harvesting and information transmission for both AF and DF networks. The idea is to adjust the time duration of energy harvesting at the relay node based on the available harvested energy and the source-to-relay channel quality. Considering the \emph{harvest-store-use} scheme at the relay node, the authors propose continuous and discrete time-switching protocols. The authors obtain analytical expressions of the achievable throughput for both the continuous and the discrete time-switching protocols. An interesting finding is that the discrete adaptive time-switching protocols, which is easier to implement, outperforms the continuous counterparts at relatively high SNR or when the SNR detection threshold is relatively low. However, the protocol only allows fixed transmit power at relay node, which may not be optimal from energy efficiency perspective. 
  
Apart from the abovementioned work \cite{IKrikidis2012,SDurrani2013,A1310.7648Nasir} which study the one-way relay network model (i.e., transmission happens in one direction), \cite{Z2013Chen} considers a two-way relay network (i.e., transmission happens in both directions) with quasi-static Rayleigh fading channels, where two source nodes exchange information through an AF-based RF-powered relay. The authors propose a power-splitting based relaying protocol and derive the exact expressions of the protocol, in terms of outage probability, ergodic capacity and finite-SNR diversity-multiplexing tradeoff. The tight closed-form lower and upper bounds of the outage probability and the ergodic capacity of the network are also obtained.   
 
Table~\ref{relay_operation} compares the reviewed relay operation policies. We observe that almost all the relay operation policies are developed for two-hop relay networks. It is also important to devise and examine operation policies for the networks with more than two hops. Moreover, the strategies to deal with co-channel interference also need to be considered in the design of the relay operation policy.   
 
\begin{table*}\footnotesize
\centering
\caption{\footnotesize Comparison of Relay Operation Policies for RFEHNs.} \label{relay_operation}
\begin{tabular}{|p{1.5cm}|p{2.8cm}|p{2cm}|p{2cm}|p{4cm}|p{2.5cm}|} 
\hline
\footnotesize {\bf Literature} & {\bf Receiver architecture} & {\bf System model} & {\bf Channel model} & {\bf Design objective} & {\bf CSI requirement} \\ \hline 
\hline 
Z. Chen \emph{et al} \cite{Z2013Chen}   & Power-splitting & Two-Way AF SISO relay system  & Quasi-static Rayleigh fading channel & Information and energy tradeoff, finite-SNR diversity-multiplexing tradeoff & None  \\  
\hline 
I. Krikidis \emph{et al} \cite{IKrikidis2012}   & Time-switching & Three-node AF SISO relay system  &  Error-free channel  & Minimization of outage probability & None    \\ 
\hline 
A.~A.~Nasir \emph{et al} \cite{SDurrani2013}   & Time-switching, power-splitting  & Three-node AF SISO relay system  &  Quasi-static block-fading channel  & Outage probability and energy tradeoff for delay-limited  transmission, information and energy tradeoff for delay-tolerant transmission &  CSI at destination node    \\ 
\hline
A.~A.~Nasir \emph{et al} \cite{A1310.7648Nasir}   & Time-switching & Three-node AF/DF SISO relay system & Quasi-static fading channels & Maximization of throughput & CSI at the receiver    \\ 
\hline
I. Krikidis \emph{et al} \cite{I2014Krikidis}   & Separated receiver architecture (antenna switching) & Three-node MIMO relay system & Rayleigh fading channel & Low complexity & None \\
 \hline
\end{tabular}
\end{table*} 
 
\subsection{Relay Selection}
  
From the network-level perspective, SWIPT gives rise to new challenges in designing the relay selection schemes for RF-EHNs. The main problem lies in that the preferable relay for information transmission does not necessarily coincide with the relay with the strongest channel for energy harvesting. Thus, as a tradeoff, relay selection has to leverage between the efficiency of information and energy transfer. 
The authors in \cite{Michalopoulos} investigate selection between two available relays in a Rayleigh fading network with a separated information receiver and energy harvester. The aim is to conduct a comparative study of three relay selection schemes, namely, time-sharing selection, threshold-checking selection and weighted difference selection scheme. In the time-sharing selection, the source node switches among the relays with the highest SNR at different time. In the threshold-checking selection, the source node chooses the relay with the highest RF energy harvesting rate. The weighted difference selection scheme selects relay based on the priority of information transmission and energy transfer. It is demonstrated that the threshold-checking selection has better performance in terms of achieved capacity for a given RF energy harvesting requirement. By contrast, the time-sharing selection has better performance in terms of outage probability when the normalized average SNR per link is larger than 5 dB. Nevertheless, both the selection methods require global CSI knowledge in each transmission session. 

The focus of \cite{I1310.6511Krikidis} is to study the impact of cooperative density and relay selection in a large-scale network with SWIPT. Specifically, the authors consider the network with a large number of randomly located transmitter-receiver pairs and potential DF relays. Both the transmitters and relays have stable power provision through wired connections, and the receivers adopt the power-splitting architecture and have both QoS and RF energy harvesting requirements. A random relay selection policy based on a sectorized area with central angle at the direction of each receiver is studied. By using the stochastic geometry model, the authors derive the closed-form function of the outage probability of the system and average harvested energy at each receiver to characterize the impact of cooperative density and relay selection area. 
  
The authors in \cite{Z1403.0354Ding} tackle the problem whether the max-min relay selection criterion, which is the diversity-optimal strategy in conventional relay network, is still diversity-optimal for relay network with RF energy harvesting. The authors consider a network with multiple source-destination pairs and one RF energy harvesting relay, where the relay schedules the user pairs for transmissions. It is found that max-min criterion will lead to the loss of diversity gains in the considered network compared to conventional network. This is because the source-relay channels and the relay-destination channels are deemed as equally in max-min criterion. However, the source-relay channels are more important than the relay-destination channels in RF-EHNs, as the former decides both the reception reliability and the harvested power at the relay. Motivated by these observations, the authors introduce a greedy scheduling algorithm which first schedules the sources with the best source-relay channel conditions, then forwards to the destinations with the best relay-destination channel conditions. It is shown that the greedy scheduling algorithm can achieve full diversity gain. However, it only works for delay tolerant networks.
  
The authors in \cite{H1404.4120Chen} devise a harvest-then-cooperate protocol, which schedules the source and relay to harvest energy first and then cooperatively performs uplink information transmission. For a single-relay scenario with delay-limited transmission, the authors derive the approximate closed-form expression for the average throughput of the proposed protocol over Rayleigh fading channels. For a multi-relay scenario, the approximate throughput functions of the proposed protocol with two relay selection schemes are derived. The simulations show that the proposed protocol is superior to the harvest-then-transmit protocol \cite{H1304.7886Ju} reviewed above in all considered cases.
   
\subsection{Power Allocation}   
 
Furthermore, research efforts attempt to address the power allocation problem in cooperative relay networks. 
In \cite{Z1307.1630ding}, the authors investigate the power allocation problem in a DF cooperative network with multiple source-destination pairs and one RF energy harvesting relay. The focus is on the strategies to distribute the harvested RF energy among the relay transmission for different source-destination pairs. The authors propose a distributed auction-based power allocation scheme based on the concept of a Nash equilibrium. Moreover, two centralized allocation schemes based on the equability principle and sequential water filling principle are also studied. The theoretical analysis shows that the water-filling based scheme is optimal in terms of the outage probability for the source-destination pairs with the worst channel conditions, while the auction-based scheme can reach the performance close to optimal. The simulation results reveal that the proposed auction-based scheme achieves good tradeoff between the system performance and complexity. However, the proposed scheme requires CSI at the transmitter, which adds significant system overhead as the number of users increases. 

The authors in \cite{K2013Tutuncuoglu} consider a two-hop relay network where multiple source nodes transmit to a common destination through a relay. Both the source nodes and the relay node are equipped with RF harvesting capacity, and can opt to transfer its energy to others for improving overall sum-rate. The authors formulate a joint transmit power allocation and energy cooperation problem to maximize the network sum-rate. It is shown that this maximization problem can be decomposed into two separated sub-problems to optimize energy transfer and transmit power allocation. The optimal energy transfer policy is exposed to be an ordered node selection problem, where nodes are prioritized according to the strength of their energy and information transmission channels. The transmit power allocation problem is solved using an iterative algorithm, which reduces to a directional water-filling algorithm when there is only one source. The authors also revisit the uni-directional energy cooperation model analyzed in \cite{BGurakan2012}, and show that the directional water-filling algorithm can solve the corresponding problem.

\subsection{Other issues}
The precoder maximization problem~\cite{Chalise2013} and cooperative schemes  \cite{Moritz2013} are also studied for the relay network with RF energy harvesting. In \cite{Chalise2013}, the authors examine a two-hop MIMO relay system with two destination nodes, i.e., an energy harvester and an information receiver. The authors investigate two scenarios. The first scenario assumes perfect CSI at receivers. The second scenario assumes only the second-order statistics of CSI at the transmitter. These statistics could be, for example, covariance matrices of the channels. The tradeoff between information rate and energy for the perfect CSI scenario is analyzed by the boundary of the rate-energy region. Then, the source and relay precoders that maximize the information transmission rate while keeping the energy transfer above a certain predefined value are designed. Likewise, the tradeoff between outage probability and energy for the second scenario is characterized by the boundary of the outage-energy region. The precoder optimization problem is formulated based on the upper bound approximation of the outage probability. The simulation results reveal that spatial correlation accounts for increased energy transfer for both scenarios. However, it also leads to the increase in outage probability, thus reducing information transmission rate.

The objective of \cite{Moritz2013} is to study the effect of cooperation schemes on energy harvesting cooperative networks. Specifically, three different cooperative schemes, namely DF, nonbinary network-coding~\cite{L2012Rebelatto} and generalized nonbinary network-coding~\cite{L2012Rebelatto}, are evaluated in the system. Multiple energy harvesting sources can work as relays for each other in uplink transmission. The authors obtain a closed-form approximation of the outage probability for each cooperative scheme. Assuming perfect CSI at receivers, the approximation for the optimal energy transfer period that minimizes the outage probability is also derived. The simulation results show that the cooperative schemes with RF energy transfer not only present lower outage probability, but also achieve higher transmission rate for a large SNR range.  
 
\section{RF-powered Cognitive Radio Networks}
\begin{figure}
\centering
\includegraphics[width=0.5\textwidth]{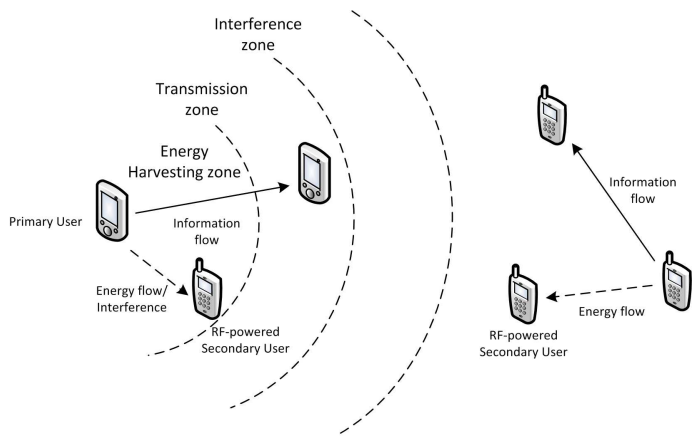}
\caption{A General Network architecture of RF-powered cognitive radio networks.} 
\label{RF-CRN}
\end{figure} 
 
Powering a cognitive radio network (CRN) with RF energy can provide a spectrum- and energy-efficient solution for wireless networking \cite{X1401.3502Lu}. 
The idea of utilizing RF signals from primary transmitters to power secondary devices has been first proposed in \cite{SLee2013}.
In an RF-powered CRN, the RF energy harvesting capability allows secondary users to harvest and store energy from nearby transmissions of primary users. Then, the secondary users can transmit data when they are sufficiently far away from primary users or when the nearby primary users are idle. Therefore, the secondary user must not only identify spectrum holes for opportunistic data transmission, but also search for occupied spectrum band/channel to harvest RF energy.

Figure~\ref{RF-CRN} shows a general network architecture for RF-powered CRNs. A secondary user can receive RF energy from a primary user on transmission. Figure~\ref{RF-CRN} also shows three zones associated with the primary user. The ``transmission zone" is the communication coverage of the primary user. Inside the ``transmission zone", if the secondary user is in the ``RF harvesting zone", the secondary user can harvest RF energy from the primary user. If the primary users occupy their channels, then the secondary user cannot transmit data if it is in the ``interference zone" (i.e., interference is created to the communication of the primary users).


\subsection{Dynamic Spectrum Access in RF-powered CRN}

Cognitive radio consists of four main functions, namely, spectrum sensing, spectrum access, spectrum management and spectrum handoff, to support intelligent and efficient dynamic spectrum access. This subsection discusses about research issues in the RF-powered CRN related to these functions.

\subsubsection{Spectrum Sensing} 
The main function of spectrum sensing in RF-powered CRNs is to detect the activities of primary users accurately.
The purpose is threefold: finding a spectrum opportunity to access for information transmission or RF energy harvesting, obtaining statistical information on spectrum usage for future reference, and predicting a potential energy level (e.g., using cyclostationary feature detection) that secondary users can harvest on a spectrum band. 

\subsubsection{Spectrum Access} 
The key issue of spectrum access is to access spectrum while protecting primary user from collision and to provide fair and efficient sharing of available spectrum.
There are two major types of spectrum access, i.e., fixed and random spectrum access. For the fixed spectrum access, radio resources are statically allocated to users. The random spectrum access allows users to contend for radio resources. These spectrum accesses are based on individual remaining energy level and available energy harvesting rate~\cite{X1401.3502Lu}.

\subsubsection{Spectrum Management}  
The objective of spectrum management in RF-powered CRNs is to achieve high spectrum utilization for both communication and RF energy harvesting by performing channel selection. In particular, for RF energy harvesting, achievable energy harvesting rate and channel occupancy  probability are the most concerned metrics in channel selection.
 
\subsubsection{Spectrum Handoff} 
Spectrum handoff is responsible for moving a secondary user from accessing one channel to another channel. In the RF-powered CRN, 
when a primary user re-occupies/releases its channel, a secondary user has to decide whether switching to another channel (if available) for information transmission/energy harvesting, or 
performing RF energy harvesting/information transmission on the re-occupied/released channel. The decision making should be made when the best time for spectrum handoff is so that the performance can be maximized.

\subsection{Review of Related Work} 
 
In \cite{S2012Park}, the authors investigate a mode selection policy for a secondary user, which casts a decision making problem between opportunistic spectrum access and RF energy harvesting. With the goal to maximize the expected total throughput, an optimal mode selection policy, balancing between the immediate throughput and harvested RF energy, is developed based on a partially observable Markov decision process. However, in \cite{S2012Park}, only a single channel in the primary network is considered. 

The authors in \cite{X1401.3502Lu} consider a channel selection policy in a multiple-channel CRN, in which the secondary user select channels not only for information transmission but also for energy harvesting. In the context of complete CSI at the secondary user, an optimal channel selection policy for the secondary user to maximize throughput is determined, based on the remaining energy level and the number of waiting packets in data queue, by applying a Markov decision process.  However, the proposed policy may impose high computation complexity on the secondary user when the state space, which is related to data queue and energy queue size, is large. 
The study in \cite{THoangJSAC} extends \cite{X1401.3502Lu} by studying the case with incomplete CSI at the secondary user. Furthermore, the authors propose an online learning algorithm for the secondary user. With the algorithm, the secondary user can use observations to adjust the channel selection strategy based on a Markov decision process to maximize throughput. Compared with the case assuming perfect CSI where throughput is optimized, the learning algorithm is shown to reach a close-to-optimal performance.

The authors in \cite{SLee2013} and \cite{A1405.2013H} analyze RF-powered CRNs with stochastic geometric approaches. The study in \cite{SLee2013} considers a network model where both RF-powered secondary users and primary users are assumed to follow independent homogeneous Poisson point processes and communicate with their intended receivers at fixed distances. The authors characterize the transmission probability of secondary users in the cases that secondary users can be fully charged within one or multiple time slots. The outage probabilities of coexisting primary and secondary networks are also derived subject to their mutual interferences. Moreover, to maximize the secondary network capacity subject to outage constraints of both primary and secondary networks, the optimal transmit power and density of secondary users are derived in closed-form. Moreover, the authors generalize the analytical results to a wireless sensor network powered by distributed wireless power chargers.

In \cite{A1405.2013H}, the authors investigate cognitive
and energy harvesting-based device-to-device (D2D) communication underlying cellular networks.
Specifically, two spectrum access policies designed for cellular base stations, namely, random spectrum access
and prioritized spectrum access, are studied. The former allows a base station to access any of the available channels randomly, while the latter let the base station access the D2D channel only when all of the other channels are occupied. Using the stochastic geometry approach, the performance of the considered system are characterized in terms of transmission probability and SINR outage probabilities for both D2D transmitters and cellular users. The simulation shows that the prioritized spectrum access method outperforms the random spectrum access method for all considered performance metrics of the D2D transmitters. Moreover, the effect of the prioritized spectrum access for the cellular users is observed to be negligible compared to the random spectrum policy.

The authors in \cite{G2014Zheng} propose a novel paradigm in RF-powered CRNs, called energy and information cooperation. The idea is that a primary network can provide both spectrum and energy to a secondary network with energy harvesting capability, so that the secondary system can assist the primary transmission in return. In this context, the authors study an ideal cooperation scheme assuming non-causal primary information available at secondary transmitters. The author then propose two schemes based on power-splitting and time-switching for SWIPT. For each scheme, both the optimal and a low-complexity solution are derived. The simulation shows that the proposed energy and information cooperation can achieve substantial performance gain compared to the conventional information cooperation only. It is also found that the power-splitting scheme can support a larger rate region than that of the time-switching scheme. 

In \cite{N.Barroca}, the authors consider a cognitive wireless body area network with RF energy harvesting capability. The authors discuss the challenges in the physical, MAC, and network layers and some potential solutions. Furthermore, practical architectures are proposed for cognitive radio-enabled RF energy harvesting devices for joint information reception and RF energy harvesting. 
 
\section{Communication Protocols}

In this section, we highlight the metrics in designing routing and MAC protocols for RF-EHNs. Additionally, the existing protocols are reviewed and compared.  

\subsection{MAC Protocol}

To achieve QoS support and fairness for information transmissions, MAC protocols designed specially for RF-EHNs are needed to coordinate the nodes' transmissions. In addition to the channel access for information transmission, the nodes also need to spend some time for RF energy harvesting. The challenge is that the time taken to harvest enough energy is different for different nodes due to various factors such as types of the available RF energy sources and distance. The MAC protocols can coordinate network nodes in a contention-free approach (e.g., polling) or a contention-based approach (e.g., CSMA/CA). The contention-free MAC protocol needs to take the node-specific RF energy harvesting process into account to achieve high throughput and fairness. With the contention-based MAC protocol, each node contends for radio resources for information transmission. If the RF energy harvesting duration is not optimally decided, an extended delay of the resource contention due to communication outage may incur.

In \cite{Kim2011}, the authors present a CSMA/CA-based energy adaptive MAC protocol for a star-topology sensor network. In the network, a single master node gathers data from and emits RF energy to a group of slave nodes. In the proposed MAC protocol, an energy adaptive duty cycle algorithm is adopted to manage the slave node's duty cycle based on the node remaining energy level. Furthermore, an energy adaptive contention algorithm is employed to use individual RF energy harvesting rate to control the corresponding backoff time. In particular, the contention algorithm compensates the unfairness caused by significant different energy harvesting rates of the slave nodes due to locations. The authors also present an analytical model to evaluate the performance of the energy adaptive MAC protocols in \cite{JKim2011}.

Nevertheless, the energy adaptive MAC protocol requires centralized control as well as out-of-band RF energy supply. It is applicable to the system with only one RF energy source. By contrast, the authors of ~\cite{Nintanavongsa2013} consider distributed control, in-band RF energy supply and multiple RF energy sources. The CSMA/CA-based MAC protocol called RF-MAC is designed to optimize RF energy delivery rate. The goal is to meet the energy requirement of sensor nodes while minimizing disruption to data communication. The RF-MAC incorporates a method to select RF energy sources to minimize the impact of interference as well as maximize energy transfer. Furthermore, the information and rate tradeoff is analyzed. Simulation results demonstrate that, compared with the modified CSMA RF-MAC, the RF-MAC is superior in terms of average harvested energy and average network throughput.

\subsection{Routing Protocol} 

For multi-hop transmission in RF-EHNs, a routing protocol that incorporates wireless charging is required to maintain end-to-end communication. In multi-hop RF-EHNs such as wireless sensor or mesh networks, as the nodes have limited internal energy reserves, they need to intelligently harvest and utilize external RF power to remain active. Therefore, being internal and external energy-aware is particularly important in the design of routing.    

\begin{figure}
\centering
\includegraphics[width=0.4\textwidth]{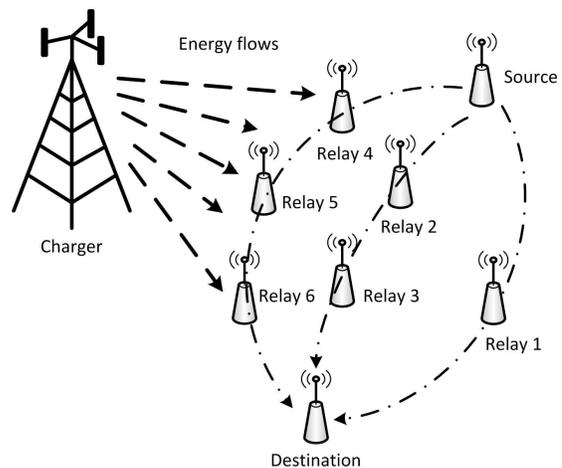}
\caption{An example of routing in RFCRN} \label{routing}
\end{figure}

Unlike the energy-aware routing developed in conventional wireless networks, the routing protocols in RF-EHNs must take the RF energy propagation and the circuit design of network nodes (e.g., RF energy harvester sensitivity) into account. This is due to the fact that the amount of harvested RF energy available at each node can be different. Moreover, the routing metric may need to be jointly defined based on energy harvesting parameters (e.g., harvester sensitivity and conversion rate, distance from RF sources, etc), spectrum management parameters (e.g., number of available channels) as well as network parameters (e.g., link quality and number of hops). Figure~\ref{routing} shows an example of the RF-EHN with an RF charger. There exist three different available routes between the source and destination represented by the dashed arrow lines. If the route with relay $1$ is selected, the delay will be smaller than that of other routes since this is the shortest route (i.e., only two hops). However, as this route is far away from the charger, the charger has to apply high power for RF energy to provide the relay $1$ with sufficient energy. By contrast, if the route with relay $4$, relay $5$ and relay $6$ is selected, these nodes near to the RF source can obtain more RF energy, and the charger does not need to use high power. However, this route may incur large delay as it is the longest route (i.e., four hops). As a tradeoff, the source may finally decide to adopt relay $2$ and relay $3$ as the intermediate nodes to leverage energy efficiency, end-to-end delay and throughput.

\subsubsection{Review of related works}

In \cite{Doost2010}, the authors consider the routing problem in a wireless sensor network where the sensor nodes are charged wirelessly over the same frequency for communication. 
It is shown by experiments that simple metrics such as hop count may not be suitable for routing in such networks. Therefore, a new routing metric based on the charging time of the sensor nodes is introduced. Then, the modified Ad hoc On-Demand Distance Vector (AODV) routing protocol considering the new routing metric is proposed. In this protocol, the sensors choose the route with the lowest value of maximum charging time. Furthermore, the link layer optimization framework is also proposed to address the tradeoff between RF energy harvesting and information transmission duration. However, this work does not address the interference problem caused to the communication by RF charging on the same frequency.  

The main concern of \cite{Y2010Peng} and \cite{L2011Zi} is to investigate how RF energy charging affects sensor network routing. The authors in \cite{Y2010Peng} first conduct experiments to study the practicability of adopting the RF charging technology to prolong the lifetime of a prototype sensor network. The considered system consists of a mobile charger with RF energy transfer capability to replenish the battery of sensors. The charger employs a simple charging strategy to replenish the sensors' battery with the lowest residual lifetime (i.e., bottleneck sensors). In this context, two well known routing protocols, i.e., energy-minimum routing and energy-balanced routing, are examined. The simulation results in a large-scale network show that the energy-balance routing achieves longer network lifetime when the charging efficiency is low or the amount of energy carried by the mobile charger is small. By contrast, when both the charging efficiency is high and the amount of energy carried by the mobile charger is large, the energy-minimum routing is superior in prolonging the network lifetime. Inspired by the observations in \cite{Y2010Peng}, the authors in \cite{L2011Zi} design a practical joint routing and charging (J-RoC) scheme for the same system model. The key idea of J-RoC is to balance energy-minimum routing and energy-balanced routing to utilize their strength as well as mitigate the shortcoming of each other. Furthermore, J-RoC requires periodic information exchange between sensors and the mobile charger. Being aware of the global energy status of the network, the mobile charger is able to schedule its charging activities. While being aware of the charging schedule, the sensors can make routing decisions based on a charging-aware routing metric. The routing metric takes into account the effects of charging activities to be executed and the real-time link quality in order to transfer the RF energy to the most demanding sensors. The simulation results show that J-RoC can approach the upper bound of network lifetime under various system configurations.
However, the scalability of J-RoC is limited as it is designed for a sensor network with a single charger.

Another work \cite{B2010Tong} focuses on the design of joint network deployment and routing strategy. The objective is to minimize the total recharging cost to enable an infinite lifetime of the network with multiple static chargers. Based on the assumption that the sensors can always be recharged in time before their power is depleted and have perfect knowledge of CSI, an optimization problem of joint network deployment and routing is formulated and proved to be NP-complete. To address this problem, the authors propose two centralized and heuristic algorithms. However, the solutions might not be practical when the link quality is imperfect or the charging capability is constrained.

\subsubsection{Comparison and discussion}

\begin{table*}
\footnotesize
\centering
\caption{\footnotesize Comparison of Routing Protocols for RFEHNs.} \label{Routing}
\begin{tabular}{|p{3cm}|l|l|p{3.5cm}|p{1.6cm}|l|l|}
\hline
\footnotesize {\bf Routing protocol} & {\bf Charging frequency} & {\bf Charger} & {\bf Route metric} & {\bf Channel state information} & {\bf Routing decision} &  {\bf Mobility} \\ \hline 
\hline
R. Doost \emph{et al}  \cite{Doost2010} & In-band & Mobile & Charging time &  High & Distributed & Limited \\
      \hline
Routing-first Heuristic Algorithm~\cite{B2010Tong} &  Out-of-band & Static & Minimum recharging cost & High & Centralized & Low \\
    \hline
Incremental Deployment-based Heuristic Algorithm~\cite{B2010Tong}  & Out-of-band & Static &  Minimum recharging cost & High & Centralized & Low \\
             \hline
J-RoC~\cite{L2011Zi} & Out-of-Band & Mobile & Charging-aware routing cost \cite{L2011Zi}, factoring the estimated energy minimum routing cost and the real-time link quality &  Medium & Distributed & Limited \\
     \hline
\end{tabular}
\end{table*}

Table~\ref{Routing} shows the comparison of the existing routing protocols for RF-EHNs.  
It can be observed that all the protocols work in the systems with a dedicated RF charger, because of which the majority of them consider out-of-band charging to avoid interference. Though in \cite{Doost2010}, the authors perform an experiment of the system where a sensor node and RF charger work on the same frequency,  no interference management scheme is taken into account. Therefore, this is a room for the investigation of routing protocols in the system adopting SWIPT. 

Furthermore, given the routing metric, information exchange among the network devices through broadcasting is required during route selection. Due to hardware limitation as aforementioned, network devices cannot harvest RF energy from the same carrier for information decoding. As a result, the network devices working in RF harvesting mode may miss the broadcast information. Thus, an efficient message broadcasting mechanism for the time-switching based receiver architecture is also required for routing protocols. %

Besides, an RF-EHN operates on ISM band (e.g., WiFi, Zigbee and Bluetooth) may overlap with the frequency band for wireless charging (e.g., the system in \cite{Doost2010}). In this context, an RF charger, if not well controlled, can cause severe interference to the network communication, as its power is usually much higher than that of network devices. Thus, there is a need for efficient spectrum allocation mechanisms to coordinate communication and charging.

\section{Future Directions and Practical Challenges}
\label{sec:openissues}
In this section, we discuss about open research issues.

\subsection{Distributed Energy Beamforming}

Distributed energy beamforming enables a cluster of distributed energy sources to cooperatively emulate an antenna array by transmitting RF energy simultaneously in the same direction to an intended energy harvester for better diversity gains. The potential energy gains at the receiver from distributed energy beamforming are expected to be the same as that from the well-known information beamforming. However, challenges arise in the implementation, e.g., time synchronization among energy sources and coordination of distributed carriers in phase and frequency so that RF signals can be combined constructively at the receiver. 


    
\subsection{Interference Management}

Existing interference management techniques, e.g., interference alignment and interference cancellation, attempt to avoid or mitigate interference through spectrum scheduling. However, with RF energy harvesting, harmful interference can be turned into useful energy through a scheduling policy. In this context, how to mitigate interference as well as facilitate energy transfer, which may be conflicting, is the problem to be addressed. Furthermore, the scheduling policy can be combined with power management schemes for further improvement in energy efficiency.

\subsection{Energy Trading}

In RF-EHNs, RF energy becomes a valuable resource. The RF energy market can be established to economically manage this energy resource jointly with radio resource. For example, wireless charging service providers may act as RF energy suppliers to meet the energy demand from network nodes. The wireless energy service providers can decide on pricing and guarantee the quality of charging service. One of the efficient approaches in this dynamic market is to develop demand side management, which allows the service providers and network nodes to interact like in smart grid, to guarantee energy-efficiency and reliability. However, the issues related to the amount of RF energy and price at which they are willing to trade while optimizing the tradeoff between the revenue and cost must be investigated.

\subsection{Effect of Mobility}

Network nodes, RF sources, and information gateway can be mobile. Therefore, mobility becomes an important factor for RF energy harvesting and information transmission. The major issue is due to the fact that the energy harvesting and information transmission performances become time-varying, and resource allocation has to be dynamic and adaptive. 


A recent work \cite{Coarasa2013} investigates the impact of mobile RF source under two different mobility models, namely \emph{center-to-center mobility (CM) model}  and \emph{around edges moving (EM) model}  with the focus on the energy gain at receivers. The tradeoff between transmit power and distance is explored, taking the energy loss during movement into account. It is found that CM yields better network performance in small networks with high node density. By contrast, EM yields better performance in large networks with low node density.

\subsection{Network Coding}

Network coding \cite{R.Ahlswede2000} is well-known to be energy efficient in information transmission. 
With network coding, senders are allowed to transmit information simultaneously. This property, especially in large-scale network, increases the amount of RF energy that can be harvested.
During the time slots when relays or senders are not transmitting, they can harvest ambient RF signals. 
A pioneer study in \cite{V.2014Mekikis} analyzes the network lifetime gain for a two-way  relay network with network coding. It is found that the lifetime of the network can be increased up to $70\%$ by enabling RF energy harvesting. From the perspective of network lifetime, more diverse network models and network coding schemes, such as physical-layer network coding and analogy network coding, are worth to be explored.  
Additionally, the energy gain of network coding has been proved to be upper bounded by $3$ in the literature. Intuitively, taking advantage of the broadcast nature of RF signals to reuse some of the dissipated energy can lead to energy saving. However, theoretically, whether RF energy harvesting will increase the upper bound of energy gain or not and how much exactly the bound will increase still require further investigation.



\subsection{Impact on Health}
 
It has long been recognized that intense RF exposure can cause heating of materials with finite conductivity, including biological tissues \cite{C.S2009Branch}. The studies in \cite{A2006W,Scientific_Committee,D2007Carpender,J2009Breckenkamp,R.W2009Habash} focus on the effects of electromagnetic waves particularly from mobile phones and cellular networks. Most of the measurements conclude that RF exposure from radio communication is safe.  However, investigations in \cite{J2009Breckenkamp} and \cite{R.W2009Habash} show that some effects to genes are noticed when the RF power reaches the upper bound of international security levels. Although there are many existing studies on the health risks of mobile phones, little effort has been made for investigation on health effect caused by a dedicated RF charger, which can release much higher power. Thus, there is a need to address the safety concerns on deploying RF chargers. 

\subsection{Practical Challenges}

\begin{itemize}

\item Due to the inverse-square law that the power density of RF waves decreases proportionally to the inverse of the square of the propagation distance, practical RF energy transfer and harvesting that complies to FCC regulations is limited to a local area. For example, the FCC allows operation up to 4W equivalent isotropically radiated power. However, as shown in \cite{T2008Le}, to realize 5.5$\mu$W energy transfer rate with a 4W power source, only the distance of 15 meters is possible. 

 
\item Other than transfer distance, RF energy harvesting rate is also largely affected by the direction and gain of the receive antenna(s). Therefore, to improve the energy harvesting efficiency, devising a high gain antenna (e.g., based on materials and geometry) for a wide range of frequency is an important research issue.

\item Impedance mismatching occurs when the input resistance and reactance of the rectifier do not equal to that of the antenna. In this context, the antenna is not able to deliver all the harvested power to the rectifier. Thus, impedance variations (e.g., introduced by on-body antennas) can severely degrade the energy conversion efficiency. There is a need to develop circuit design techniques that automatically tune the parameters to minimize impedance mismatch.


\item The RF-to-DC conversion efficiency depends on the density of harvested RF power. Improving the RF-to-DC conversion efficiency at low harvested power input is important. Moreover, realizing a high-efficient low-power DC-to-DC converter, which converts a source of DC from a voltage level to another, would be another effort to achieve highly efficient RF energy harvesting.

\item RF energy harvesting components need to be small enough to be embedded in low-power devices. For example, the size of an RF-powered sensor should be smaller than or comparable to that of a battery-power sensor. As introduced above, an RF energy harvesting component may require an independent antenna, matching network and rectifier. The antenna size has a crucial impact on an energy harvesting rate. Additionally, high voltage at the output of a rectifier requires very high impedance loads (e.g., 5M$\Omega$), which is a function of the length of the impedance. Thus, it is challenging to reduce the size of embedded devices while maintaining high energy harvesting efficiency.

\item Without line-of-sight for RF waves from an RF source to an energy harvester, the considerable energy transfer loss is expected. Therefore, the RF energy source must be optimally placed to support multiple receivers to be charged. Moreover, in a mobile environment, the mobility of receivers and energy sources can affect the RF energy transfer significantly.

\item The sensitivity of an information receiver is typically much higher than that of an RF energy harvester. Consequently, a receiver located at a distance away from an RF transmitter may be able only to decode information and fail to extract energy from the RF signals. In this case, any SWIPT scheme cannot be used efficiently. Therefore, improving the sensitivity of RF energy harvesting circuit is crucial. 

\item For RF-powered devices, as the transmit power is typical low, multiple antennas can be adopted to improve the transmission efficiency. However, larger power consumption comes along when the number of antennas increases. Thus, there exists a tradeoff between the transmission efficiency and power consumption. The scheme to optimize this tradeoff needs to be developed. This issue becomes more complicated in a dynamic environment, e.g., with varying energy harvesting rate.

\item As RF-powered devices typically have a strict operation power constraint, it is not practical to support high computation algorithms. Any schemes, such as modulation and coding, receiver operation policy and routing protocol, to be adopted need to be energy-efficient and low-power. Hence, power consumption is always a serious concern in RF-powered devices, which may require the re-design of existing schemes and algorithms for conventional networks.


\end{itemize}

\section{Conclusion}

We have presented a comprehensive survey on RF energy harvesting networks (RF-EHNs). Firstly, we have provided an overview of RF-EHNs with the focus on architecture, enabling techniques and existing applications. Then, we have reviewed the background in circuit design and state-of-the-art circuitry implementations. Afterwards, we have surveyed various design issues related to resource allocation in RF-EHNs, and the up-to-date solutions. Finally, we have discussed on the future directions and practical challenges in RF energy harvesting techniques.

\end{document}